\documentclass{aa}

\usepackage{amsmath}
\makeatletter

\pdfpageheight\paperheight
\pdfpagewidth\paperwidth

\newcommand{\lyxmathsym}[1]{\ifmmode\begingroup\def\b@ld{bold}
  \text{\ifx\math@version\b@ld\bfseries\fi#1}\endgroup\else#1\fi}

\newcommand{\lowell}{<\ell}
\newcommand{\bigell}{\geqslant\ell}

\numberwithin{equation}{section}
\numberwithin{figure}{section}

\makeatother

\usepackage{babel}

\usepackage[dvipsnames,table,xcdraw]{xcolor}
\newcommand{\tcg}[1]{\textcolor{black}{#1}}

\begin{document}

\title{Multiscale Turbulence Synthesis: validation in 2D hydrodynamics}
\author{P. Lesaffre\inst{\ref{inst:LPENS},\ref{inst:Obs}}, 
J.-B. Durrive\inst{\ref{inst:IRAP}}, 
J. Goossaert\inst{\ref{inst:LPENS}}, 
S. Poirier\inst{\ref{inst:LPENS}}, 
S. Colombi\inst{\ref{inst:IAP}},
P. Richard\inst{\ref{inst:LPENS},\ref{inst:Obs}},
E. Allys\inst{\ref{inst:LPENS},\ref{inst:Obs}},
W. B\'ethune\inst{\ref{inst:ONERA}}
}

\institute{Laboratoire de Physique de l'\'Ecole Normale Sup\'erieure, ENS, Universit\'e PSL, CNRS, Sorbonne Universit\'e, Universit\'e Paris Cit\'e, F-75005, Paris, France \label{inst:LPENS}
\and
 Observatoire de Paris, PSL University, Sorbonne Universit\'e, LERMA, 75014, Paris, France \label{inst:Obs}
\and
Institut de recherche en astrophysique et plan\'etologie Universit\' Toulouse III - Paul Sabatier, Observatoire Midi-Pyr\'en\'ees,
Centre National de la Recherche Scientifique, UMR5277 France \label{inst:IRAP}
\and 
Sorbonne Université, CNRS, UMR7095, Institut d’Astrophysique de Paris, 98 bis Boulevard Arago, 75014 Paris, France \label{inst:IAP}
\and
DPHY, ONERA, Université Paris-Saclay, 91120, Palaiseau, France \label{inst:ONERA}
}
\titlerunning{MuScaTS}
\date{ ?? / ??}

\abstract
{Numerical simulations can follow the evolution of fluid motions through the intricacies of developed turbulence. 
However, they are rather costly to run, especially in 3D. In the past two decades, generative models have emerged which produce synthetic random flows at a computational cost equivalent to no more than a few time-steps of a simulation.
These simplified models qualitatively bear some characteristics of turbulent flows in specific contexts (incompressible 3D hydrodynamics or magnetohydrodynamics), but generally struggle with the synthesis of coherent structures.}
{We aim at generating random fields (e.g. velocity, density, magnetic fields, etc.) with realistic physical properties for a large variety of governing partial differential equations and at a small cost relative to time-resolved simulations.}
{We propose a set of simple steps applied to given sets of partial differential equations: filtering from large to small scales, first order time evolution approximation from Gaussian random initial conditions during a prescribed coherence time, with a final summation over scales to generate the whole fields.}
{
We test the validity of our method in the simplest framework: 2D decaying incompressible hydrodynamical turbulence. We compare results of 2D decaying simulations with snapshots of our synthetic turbulence. We assess quantitatively the difference first with standard statistical tools: power spectra, increments and structure functions. These indicators can be reproduced by our method during up to about a third of the turnover time scale. We also consider recently developed scattering transforms statistics, able to efficiently characterise non-Gaussian structures. This reveals more significant discrepancy, which can however be bridged by bootstrapping. Finally, the number of Fourier transforms necessary for one synthesis scales logarithmically in the resolution, compared to linearly for time-resolved simulations.
}
{We have designed a multiscale turbulence synthesis (MuScaTS) method to efficiently short-circuit costly numerical simulations to produce realistic instantaneous fields.}

\maketitle

\section{Introduction and context}

Turbulence has distinctive features, which are hard to characterise quantitatively.
Nevertheless, our subjective experience tells us that a lot of information is encoded even in a single 2D projection of a tracer in a 3D turbulent flow. Indeed, when we give a vigorous push to a puff of smoke, it soon develops convoluted motions, diverging trajectories blur the initial impulse, but nevertheless some characteristic signatures remain. When we look at a still picture of smoke trailing in the wake of a candle or an incense stick, our brain not only immediately recognises the context, it also grasps a sense for the fluid shearing and spiralling motions in vortices and gets a feeling of the large scale wind direction.  Astronomers who study the interstellar medium are very much in the same situation: the sky before their eyes is a still projection of tracers evolving through the fluid dynamics at play in these dilute media, which they attempt to infer. 

Quantitative probes of the nature of the underlying fluid dynamics are however quite elusive. Turbulence research so far has managed to isolate only a few generic statistical properties shared by turbulent flows. In the case of incompressible hydrodynamics (HD), power-law scalings for the velocity power spectrum were first obtained by \cite{K41} using conservation arguments across scales. But scaling laws for the velocity and magnetic fields power spectra are still a matter of debate even in the relatively simple case of incompressible magnetohydrodynamic (MHD) turbulence \citep{Iroshnikov1963,Kraichnan1965,Goldreich1995,Schekochihin2022}. 
Then, anomalous scalings for the structure functions were discovered \citep{K62}, which led to a large legacy of intermittency models and observations of statistical properties on the increments of velocity, energy dissipation rates, vorticity or magnetic fields \citep{Frisch,She1994,Grauer1994,Politano1995}. 
Finally, few exact analytical results were predicted such as the K\'ar\'man-Howarth-Monin 4/5th relation \citep{Karman1938,Monin1959,Frisch} in 3D incompressible HD and its analogues in compressible or MHD turbulence \citep{Galtier2011,Banerjee2013,Banerjee2014,Banerjee2016}.

In contrast, time-resolved simulations can accurately predict the evolution of fluid motions from random initial conditions and we could try to assess what statistical laws control their outcomes. But such simulations are rather costly to run, especially in 3D and in a developed turbulent regime, which limits the number of possible independent realisations for these experiments. 

In the past two decades, several authors have developed efficient techniques to produce random fields that bear some of the known characteristics of turbulence. 
A natural starting step is simply to produce a Gaussian velocity field with a prescribed power spectrum \citep{Mandelbrot1968}. Application of a non-linear transformation to this fractional Gaussian field can then yield some degree of non-Gaussianity, such as intermittency. However, these non-Gaussianities may have nothing to do with actual physics. For instance, it is challenging to generate asymmetries in the longitudinal velocity increments which generate non-zero energy transfer, let alone transfer in the right scale direction.

Some techniques focus on advection. \citet{Rosales2006} (multiscale minimal Lagrangian map, MMLM) and \citet{2008PhRvE..78a6313R} (multiscale turnover minimal Lagrangian map, MTML) transport ballistically an initial fractional Gaussian velocity field under its own influence in a hierarchy of embedded smoothing scales. They show this procedure generates anomalous scalings and energy transfer although they do not compare it quantitatively to experiments. \citet{Subedi2014} later extended the same approach in MHD. More recently \citet{Luebke2024} showed they could initiate the formation of some current sheets if they applied the same technique to an already non-Gaussian initial field. 

Other techniques focus on deformation, neglecting advection. \cite{2006PhRvL..97q4501C} and \cite{Chevillard2011} deform a Gaussian field with a stretching matrix exponential to produce a tunable random field with given power spectrum and intermittency properties. \cite{Durrive2020} later generalised the same ideas to incompressible MHD. 
However, these models fail at reproducing coherent structures seen in direct numerical simulations (DNS, see \citealt{Momferratos2014,2022A&A...664A.193R}), so \cite{Durrive2022} extended their model to include prescribed random dissipative sheets. 

In this paper, we present a set of generic ideas that allow to produce synthetic models for many types of turbulent flows (HD incompressible or isothermal, self-gravitating or not, incompressible MHD, etc...). We present our multiscale turbulence synthesis (MuScaTS) technique in section 2 and compare it to previous models in section 3. In section 4, we carefully assess its validity in the simplest framework of 2D incompressible hydrodynamical turbulence. To this aim, on one hand we compute the results of DNS evolving an initial Gaussian field after some finite time, and, on the other hand, we generate snapshots of synthetic turbulence using several variants of our method. In order to quantitatively compare the two, we compute power spectra, increments statistics, transfer functions and we finally assess the quality of the textures we generate by computing scattering transforms statistics (see \citealp{2019A&A...629A.115A}).  We discuss our results in section 5 and develop the prospects of the MuScaTS method in section 6. 

\section{Synthetic models of turbulence: the MuScaTS technique}
In the following subsections, we present the ingredients of our framework: a multiscale filtering \eqref{sec:filtering}, and a generic form \eqref{sec:generic-form} for partial differential equations which survives filtering under some additional approximations \eqref{sec:derivation}. This allows us to build a synthetic flow with a sweep from large to small scales \eqref{sec:synthesis}. We then discuss in more details our two main approximations regarding time evolution \eqref{subsec:Time-evolution} and coherence time estimation \eqref{subsec:coherence-time}.

\subsection{Isotropic filtering}
\label{sec:filtering}

The purpose of this section is to define a continuous series of filters labelled by their respective scales. In particular, this will allow us to reconstruct any zero mean field from its filtered components, and to separate it into a small and a large scale components.

We consider tensor fields (such as the density or the velocity fields, for example) on a space domain ${\cal D} = \mathbb{R}^{d}$ of dimension $d$.  For a field $f$, we define its filtered version $f_\ell$ at a scale $\ell$ through:
\begin{equation}
\label{eq:filtering}
f_{\ell}=\varphi_{\ell}*f,
\end{equation}
where the sign $*$ denotes convolution and $\varphi_{\ell}$ is a bandpass filter which selects scales on the order of $\ell$.

We define a continuous collection of filters labelled by all possible scales $\ell\in\mathbb{R}^+$ by dilation from a single mother filter $\varphi$ through the formula 
\begin{equation}
\label{eq:dilation-x}
    \varphi_{\ell}(\boldsymbol{x})=\varphi(\boldsymbol{x}/\ell)/\ell^{d},
\end{equation} 
 where the mother filter $\varphi$ selects scales on the order of 1, i.e. its Fourier transform $\hat{\varphi}(\boldsymbol{k})$ is real positive, decays at small and large wavenumbers and peaks at around $|\boldsymbol{k}|=1$, where $|\boldsymbol{k}|$ denotes the norm of $\boldsymbol{k}$. Note that since $\hat{\varphi}(\boldsymbol{0})=0$, $\varphi$ has zero mean and so has $f_\ell$.
 The dilation formula in Fourier space then takes the form

\begin{equation}
\label{eq:dilation-k}
    \hat{\varphi}_{\ell}(\boldsymbol{k})=\hat{\varphi}(\ell \boldsymbol{k}).
\end{equation}

We further assume {\it isotropic} filtering, i.e. $\hat{\varphi}(\boldsymbol{k})$ depends only on $|\boldsymbol{k}|$ through the function $\Phi(s)$ defined for $s\in\mathbb{R}^+$ such that 
\begin{equation}
\label{eq:isotropic_filter}
    \hat{\varphi}(\boldsymbol{k})=\Phi(|\boldsymbol{k}|).
\end{equation}

Now, let us define
\begin{equation}
\label{eq:normalisation-constant_def1}
        C_\Phi \equiv \int_{0}^{+\infty} \frac{{\rm d} s}{s}\,\Phi(s),
\end{equation}
or equivalently, using a short-hand notation that we will use throughout the paper for such integrals,
\begin{equation}
\label{eq:normalisation-constant_def2}
        C_\Phi \equiv \int_{-\infty}^{+\infty}{\rm d}\ln s\,\Phi(s).
\end{equation}
We will assume that this integral is finite $(0<C_\Phi<+\infty)$, which implicitly constrains the shape of our mother filter $\hat{\varphi}$ because $\hat{\varphi}$ should decay fast enough at both $|\boldsymbol{k}|\rightarrow0$ and $|\boldsymbol{k}|\rightarrow +\infty$ in order to keep this integral finite.
Without loss of generality, we can assume that 
\begin{equation}
\label{eq:normalisation}
C_\Phi=1,
\end{equation}
 by redefining $\Phi \rightarrow \Phi/C_\Phi$.
 For example, the choice 
\begin{equation}
\Phi(s) \equiv s^2e^{-s^2}/\int_{0}^{+\infty}x e^{-x^2}{\rm d} x,
\end{equation}
results in a normalised $C_\Phi$.

In appendix \ref{sec:reconstruction_formula_demo} we show that the normalisation \eqref{eq:normalisation} allows us to reconstruct simply any field $f$ (with zero mean) from its filtered components $f_{\ell}$, namely we have the following reconstruction formula

\begin{equation}
\label{eq:reconstruction}
f=\int_{-\infty}^{+\infty}{\rm d}\ln\ell\,f_{\ell}.
\end{equation}

Finally, we denote 
\begin{equation}
\label{eq:lowell}
    f_{\lowell}=\int_{s\lowell}{\rm d}\ln s\,f_{s}
\end{equation}
and 
\begin{equation}
\label{eq:bigell}
    f_{\bigell}=\int_{s\bigell}{\rm d}\ln s\,f_{s}
\end{equation} 
the components of $f$ at scales respectively smaller and larger than a given $\ell$, so that 
\begin{equation}
\label{scale_separation}
    f=f_{\lowell}+f_{\bigell} .
\end{equation}

\subsection{Scope of our method: generic evolution equations}
\label{sec:generic-form}

We present here a generic form of the partial differential equations for which the approximations inherent to our method can be more easily justified. 
Consider a $n$-components vector $\boldsymbol{W}$ characterising the state of the gas as a field on our space domain ${\cal D}$. 
For example, for incompressible 3D fluid dynamics, we can take $\boldsymbol{W}=(\boldsymbol{w})$, with $\boldsymbol{w}$ the vorticity vector. Or we could use $\boldsymbol{W}=(\boldsymbol{w},\rho)$ with $\rho$ the mass density for compressible HD. 

We will then consider evolution equations in the form
\begin{equation}
\label{eq:generic-form}
\partial_{t}\boldsymbol{W}+\boldsymbol{(v[W].\nabla)W}=\boldsymbol{S[W].W}+\boldsymbol{D[W]},
\end{equation}
where $\boldsymbol{v}$ is an advection velocity vector (i.e. it has the same number of components $d$ as the space dimension), $\boldsymbol{S}$ is a $n\times n$ deformation matrix and $\boldsymbol{D}$ is a term with $n$ components which can (generally but not always) characterise the diffusion or dissipation terms. Here, the brackets $\text{\ensuremath{\boldsymbol{[W]}}}$ indicate that $\boldsymbol{v},\boldsymbol{S}$ and $\boldsymbol{D}$ are affine functions of the state variables field $\boldsymbol{W}$. In other words, we require  the non-linearities to be at most quadratic in the state vector $\boldsymbol{W}$.
\tcg{We keep the freedom of $\boldsymbol{v}$ having a linear dependence on $\boldsymbol{W}$, to accommodate future cases where the advection velocity might be more complicated, but all practical results presented in this paper are for $\boldsymbol{v}[\boldsymbol{W}] = \boldsymbol{u}$, where $\boldsymbol{u}$ is the velocity component in the state vector $\boldsymbol{W}$.}

In the following paragraphs, we give a few explicit examples of evolution equations which fit our framework, for the vorticity and divorticity in incompressible 2D hydrodynamics, and for the vorticity in 3D incompressible hydrodynamics. We then list a few more examples without giving proof.

\subsubsection{Incompressible 2D hydrodynamics}
As a first example, consider the evolution equation for 2D incompressible hydrodynamics, which reads:
\begin{equation}
\label{eq:incompressible-2D}
\partial_{t}w+\boldsymbol{\boldsymbol{u.\nabla}}w=\nu\Delta w,
\end{equation}
where $w$ is the vorticity component along the normal
to the space domain, $\boldsymbol{u}$ is the fluid velocity and $\nu$
is the kinematic viscosity coefficient. It takes the form \eqref{eq:generic-form} if we simply set $\boldsymbol{W}=(w)$, $\boldsymbol{v[W]=u}$, the
fluid velocity as expressed from $w$ by the 2D Biot-Savart
formula (without the boundaries' term)
\begin{equation}
\label{eq:Biot-Savart-2D}
    \boldsymbol{u}(\boldsymbol{x})=\int_{\cal{D}}{\rm d}^2{y}\,w(\boldsymbol{y}) \boldsymbol{\hat{z}}\times (\boldsymbol{x-y})/|\boldsymbol{x-y}|^{2}/(2\pi), 
\end{equation}
where $\boldsymbol{\hat{z}}$ is the unit vector normal to the space domain $\cal{D}$,
$\boldsymbol{S}=\boldsymbol{0}$ and $\boldsymbol{D[W]}=\nu\Delta w$. Note that equation \eqref{eq:Biot-Savart-2D} can be efficiently computed by using Fourier transforms.

\subsubsection{Divorticity in 2D hydrodynamics}

\label{sec:generic-biv}
In 2D incompressible hydrodynamics, it turns out one can also write evolution equations for the curl of the vorticity, a.k.a. the divorticity \citep{Kida1985,Shivamoggi2024}:
\begin{equation}
\label{eq:divorticity}
\boldsymbol{B}\equiv \boldsymbol{\nabla}\times\left(w\boldsymbol{\hat{z}}\right).
\end{equation}

Taking the curl of the evolution equation of vorticity \eqref{eq:incompressible-2D}, it results:
\begin{equation}
\partial_{t}\boldsymbol{B}+\boldsymbol{u.\nabla B}=\boldsymbol{S.}\boldsymbol{B}+\nu\Delta\boldsymbol{B},\label{eq:divorticity-evolution}
\end{equation}
where $\boldsymbol{S}$ is the velocity gradient matrix, so that $\boldsymbol{S.}\boldsymbol{B}=\boldsymbol{u.\nabla B}$. 
This equation is in the form \eqref{eq:generic-form} when we set $\boldsymbol{W}=(\boldsymbol{B})$, and it is also strikingly similar to the equation for the evolution of vorticity in 3D, see \eqref{eq:incompressible-3D} below. Note that the velocity field can be recovered from the divorticity by inverting the curl twice, which is a linear operation, and so the matrix $\boldsymbol{S}$ also depends linearly on the field $\boldsymbol{B}$. We will use equation \eqref{eq:divorticity-evolution} as a test bed in 2D on how to handle the respective advection and deformation terms.

\subsubsection{3D incompressible hydrodynamics}
For 3D incompressible hydrodynamics, the evolution equation for the
vorticity $\boldsymbol{w}$ reads 
\begin{equation}
\label{eq:incompressible-3D}
\partial_{t}\boldsymbol{w}+\boldsymbol{u.\nabla}\boldsymbol{w}=\boldsymbol{w.\nabla}\boldsymbol{u}+\nu\Delta\boldsymbol{w}.
\end{equation}

To show that this equation takes our generic form \eqref{eq:generic-form}, we can set $\boldsymbol{W}=\boldsymbol{w} \equiv \boldsymbol{\nabla\times u}$
the vorticity vector, and then \tcg{(again, $\boldsymbol{v}[\boldsymbol{W}] = \boldsymbol{u}$ as explained in the introduction of this section \ref{sec:generic-form})}
\begin{equation}
\label{eq:Biot-Savart-3D}
\boldsymbol{u}=
\int_{{\cal D}}{\rm d}^3{y}\,\boldsymbol{w}(\boldsymbol{y}) \times (\boldsymbol{x-y})/|\boldsymbol{x-y}|^{3}/(4\pi)
\end{equation}
(3D Biot-Savart formula without the boundaries' term). $\boldsymbol{S}$
is the velocity gradient matrix and $\boldsymbol{D[W]}=\nu\Delta\boldsymbol{w}$ is simply the diffusion term. 

Note that if we write the incompressible Navier-Stokes equation for the velocity evolution, with the specific pressure
$p \equiv P/\rho$ (with $P$ the thermal pressure and $\rho$ the mass density):
\begin{equation}
\label{eq:incompressible-3D-P}
\partial_{t}\boldsymbol{u}+
\boldsymbol{u.\nabla}\boldsymbol{\boldsymbol{u}}=
-\boldsymbol{\boldsymbol{\nabla}}p+\nu\Delta\boldsymbol{u},
\end{equation}
it does not take the form \eqref{eq:generic-form}. Indeed, for the
choice $\boldsymbol{W=u}$, the specific pressure field and its gradient
can classically be recovered in the Leray formulation \cite[][section 1.8]{Majda_Bertozzi_2001} by solving for $p$ in $\Delta p=-\partial_{i}u_{j}.\partial_{j}u_{i}$, but $\boldsymbol{\nabla} p$ then cannot seem to be put into the form $\boldsymbol{S[u].u}$ with $\boldsymbol{S[u]}$ affine in $\boldsymbol{u}$. However, this is not a restriction since we can simply use equations \eqref{eq:incompressible-3D} with \eqref{eq:Biot-Savart-3D}, but we feel this example clarifies the subtleties in the scope of the generic form \eqref{eq:generic-form}.

\subsubsection{Other cases}
\label{sec:scope-others}
The evolution equations for reduced MHD, incompressible MHD, or self-gravitating isothermal gases can also be massaged into our generic form \eqref{eq:generic-form}.
However, we could not make compressible MHD strictly fit our framework, as the Lorentz term combines a product of three primitive variables (and our attempts at changing variables always led to a product of three variables). 

In the context of large scale structure formation in cosmology, Euler-Poisson equations in the dust approximation (zero pressure) are easily cast into our generic form \eqref{eq:generic-form} for the state vector $\boldsymbol{W}=(\boldsymbol{u},\rho)$. Indeed, Poisson equation for the gravitational potential is linear in the density. The expansion of the Universe can also be incorporated by switching to properly defined comoving variables. We will use this in section \ref{sec:Zeldo} to create a link with the multiscale Zel'dovich approximation \citep{BondMyers1996,MonacoTTGQS2002,SteinAlvarezBond2019}.
Nevertheless, the form \eqref{eq:generic-form} allows for a large variety of fluid evolution equations. 

\subsection{Derivation of the filtered evolution equations}
\label{sec:derivation}

\begin{figure}
    \centering
    \includegraphics[width=\columnwidth]{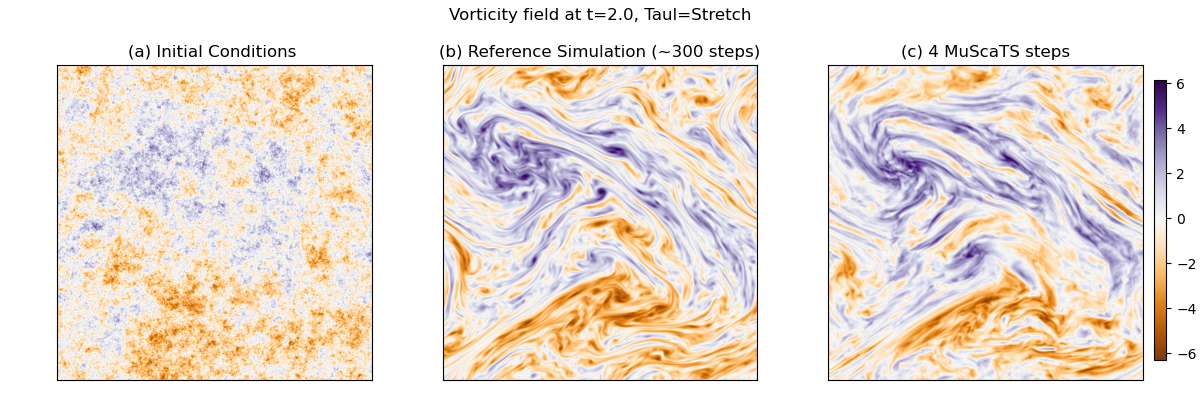}
    \caption{Example of the performance of our method. Panel (a) shows the common initial (Gaussian) vorticity field used in the two other panels, namely a standard 2D incompressible hydrodynamics simulation (panel (b)) and one of our synthesis (panel (c)). Panel (b) shows the vorticity field after the field has been evolved for about one third of a turnover time, using more than 300 standard simulation steps. Panel (c) shows the resulting synthesis after only 4 MuScaTS steps for a total computer processing unit (CPU) cost 17 times smaller.}
    \label{fig:teaser}
\end{figure}

In this section we present a series of approximations leading to a simplified description of the dynamics, namely equation \eqref{eq:Wtildel-vtilde}. From section \ref{sec:synthesis} and onward, the latter equation will be the central focus, as, instead of trying to time integrate the more complicated full partial differential equation \eqref{eq:generic-form}, we will mimic the dynamics from \eqref{eq:Wtildel-vtilde}. Thus, we will be able to generate cheap but realistic turbulent fields, as illustrated in figure \ref{fig:teaser}.

As a first step, let us recast the evolution equation \eqref{eq:generic-form} into the form \eqref{eq:filtered_EoM_2} below, which spells out how the various scales contribute to the dynamics. To do so, note that because our filtering \eqref{eq:filtering} is a convolution product, it commutes with linear functions\footnote{For affine functions, the affine constant becomes filtered, so the notations have to be slightly amended, but the derivation is the same.}, so we have in particular
\begin{equation}
\begin{array}{l}
\boldsymbol{v[W}]_{\ell}=\boldsymbol{v[W}_{\ell}\boldsymbol{]}, \\
\boldsymbol{S[W}]_{\ell}=\boldsymbol{S[W}_{\ell}\boldsymbol{]}, \\
\boldsymbol{D[W}]_{\ell}=\boldsymbol{D[W}_{\ell}\boldsymbol{]},
\end{array}
\label{eq:linearity}
\end{equation}
and, filtering \eqref{eq:generic-form} with a given bandpass filter $\varphi_\ell$,
\begin{equation}
\label{eq:filtered_EoM_1}
\partial_{t} \boldsymbol{W}_{\ell} =  
\varphi_\ell*(-\boldsymbol{v[W].\nabla W}+\boldsymbol{S[W].W})+\boldsymbol{D[W}_\ell\boldsymbol{]}.
\end{equation}
Then, using the scale decomposition \eqref{scale_separation}, we separate the large and small-scale parts of the velocity field as
\begin{equation}
\boldsymbol{v}
=
\boldsymbol{v}[\boldsymbol{W}_{\lowell}+\boldsymbol{W}_{\bigell}]
=
\boldsymbol{v}[\boldsymbol{W}_{\lowell}]
+
\boldsymbol{v}[\boldsymbol{W}_{\bigell}]
=
\boldsymbol{v}_{\lowell}
+
\boldsymbol{v}_{\bigell},
\end{equation}
and doing the same for $\boldsymbol{S}$ we rewrite the filtered evolution equation \eqref{eq:filtered_EoM_1} as
\begin{multline}
\label{eq:filtered_EoM_2}
\partial_{t}\boldsymbol{W}_{\ell}\rfloor_{\boldsymbol{x}}
= \boldsymbol{D}[\boldsymbol{W}_{\ell}] \rfloor_{\boldsymbol{x}}\\
-\int_{{\cal D}}{\rm d}^d{y}\,
\varphi_{\ell}\rfloor\boldsymbol{_{y}}
\boldsymbol{v}_{\bigell}\boldsymbol{.\nabla W}\rfloor{}_{\boldsymbol{x}-\boldsymbol{y}}
-\int_{{\cal D}}{\rm d}^d{y}\,
\varphi_{\ell}\rfloor\boldsymbol{_{y}}
\boldsymbol{v}_{\lowell}\boldsymbol{.\nabla W}\rfloor{}_{\boldsymbol{x}-\boldsymbol{y}} \\
+\int_{{\cal D}}{\rm d}^d{y}\,
\varphi_{\ell}\rfloor\boldsymbol{_{y}}
\boldsymbol{S}_{\bigell}\boldsymbol{.W}\rfloor{}_{\boldsymbol{x}-\boldsymbol{y}}
+\int_{{\cal D}}{\rm d}^d{y}\,
\varphi_{\ell}\rfloor\boldsymbol{_{y}}
\boldsymbol{S}_{\lowell}\boldsymbol{.W}\rfloor{}_{\boldsymbol{x}-\boldsymbol{y}},
\end{multline}
where ${\cal D}$ is the space domain. We expressed convolution products with a notation of the form $\rfloor_{\cdots}$ to help the reader keep track of which variable each field depends on. In particular bear in mind that \eqref{eq:filtered_EoM_2} is evaluated at position~$\boldsymbol{x}$.

The above lengthy but explicit formulation of the dynamics is the viewpoint from which we now proceed to a series of simplifying assumptions, which will result in the approximate evolution equation \eqref{eq:Wtildel-vtilde}.

Our first approximation consists in considering that the factors $\boldsymbol{v}_{\bigell}$ and $\boldsymbol{S_{\bigell}}$ appearing inside two of the integrals in \eqref{eq:filtered_EoM_2} are slowly varying in space compared to the filter $\varphi_{\ell}$, which can be understood as a scale separation approximation. Since these factors presumably remain approximately constant over the local kernel of $\varphi_{\ell}$, we pull them out of the integrals and write
\begin{multline}
\label{eq:full-filtered}
\partial_{t}\boldsymbol{W}_{\ell}
\simeq 
-\boldsymbol{v}_{\bigell} \boldsymbol{.\nabla W}_{\ell} 
+\boldsymbol{S}_{\bigell} \boldsymbol{.W}_{\ell}
+\boldsymbol{D}[\boldsymbol{W}_{\ell}] \\
-\int_{{\cal D}}{\rm d}^d{y}\,
\varphi_{\ell}\rfloor\boldsymbol{_{y}}
\boldsymbol{v}_{\lowell}\boldsymbol{.\nabla W}\rfloor{}_{\boldsymbol{x}-\boldsymbol{y}}
+\int_{{\cal D}}{\rm d}^d{y}\,
\varphi_{\ell}\rfloor\boldsymbol{_{y}}\boldsymbol{S}_{\lowell}\boldsymbol{.W}\rfloor{}_{\boldsymbol{x}-\boldsymbol{y}}.
\end{multline}
In so doing, we have also taken the $\boldsymbol{\nabla}$ sign out of the integrals, an operation allowed by the property of the gradient of a convolution product $\partial_i(f*g) = f*\partial_i g$ for multivariable functions $f$ and $g$. The last two terms on the right hand side are solely due to scales $\ell$ and below, and represent respectively advection and deformation from small scales convolved with the filter at scale $\ell$.

As a second approximation, we consider that these last two terms evolve more rapidly in time than the previous large scale terms, and somewhat chaotically, i.e. scales smaller than $\ell$ introduce some stochastic perturbations which affect the scale $\ell$. Schematically:
\begin{multline}
\partial_{t}\boldsymbol{W}_{\ell} \simeq
-\boldsymbol{v}_{\bigell}\boldsymbol{.\nabla W}_{\ell}
+\boldsymbol{S}_{\bigell}\boldsymbol{.W}_{\ell}+\boldsymbol{D[W}_{\ell}\boldsymbol{]} \\
+\textrm{"smaller scales stochasticity"}.
\end{multline}
Nevertheless, we expect the evolution to remain coherent over some time $\tau_{\ell}$ at a given scale $\ell$. Thereafter, we will either prescribe or estimate this time scale $\tau_{\ell}$ from the larger scale fields, and we will provide several ways to estimate it in section \ref{subsec:Time-evolution} below. 
For now, since $\tau_{\ell}$ controls the time scale over which two initially adjacent fluid particles diverge from each other, it can be seen as the inverse of the largest positive Lyapunov exponent under the combined effects of the advection flow $\boldsymbol{v}_{\bigell}$ and the deformation matrix $\boldsymbol{S}_{\bigell}$. 

This sensitivity to the initial conditions implies that it is not quite necessary to integrate precisely the flow equations during more than a coherence time. Indeed, if the conditions had been slightly different one coherence time before, the current state of the fluid might now be completely different, to a point these initial conditions might as well be chosen randomly. Therefore, rather than solving a complicated stochastic equation, we will also assume that the effect of the chaos in the small scales is equivalent to randomly reshuffle initial conditions every coherence time $\tau_{\ell}$. 

Thanks to this hypothesis, we do not need to integrate the evolution from the beginning, but only during the length of the last coherent time interval.
At each scale $\ell$, we hence assume that the chaotic evolution can be approximated by deterministically integrating to the targeted time $t$ from random conditions at time $t-\tau_{\ell}$ during a coherence time interval $\tau_{\ell}$ . We denote this change by adding tildes over the vectors subject to this approximation. 
As a result, we approximate the evolution of the field as:
\begin{equation}
\partial_{t}\boldsymbol{\tilde{W}}_{\ell}=
-\boldsymbol{v}_{\bigell}\boldsymbol{.\nabla}\boldsymbol{\tilde{W}}_{\ell}+
\boldsymbol{S}_{\bigell}\boldsymbol{\tilde{W.}}_{\ell}+
\boldsymbol{D[\tilde{W}}_{\ell}\boldsymbol{]},\label{eq:Wtildel}
\end{equation}
starting from random  conditions $\boldsymbol{W}_{\ell}^{0}$  at intermediate time $t_{\ell}^{0}=t-\tau_{\ell}$ (these intermediate times might hence depend on scale). 

The law of $\boldsymbol{W}_{\ell}^{0}$ still needs to be specified. The chaos induced by the small scales might well be non trivial, but in a first step we are going to assume that it follows a Gaussian law. Hence we assume that $\boldsymbol{W}_{\ell}^{0}$ is a filtered version $\boldsymbol{W}_{\ell}^{0}=\varphi_{\ell}*\boldsymbol{W}^{0}$ of a Gaussian field $\boldsymbol{W}^{0}$, for which the spectrum needs to be prescribed. 
Furthermore, to reflect the property that neighbouring scales are in fact not independent (we expect the behaviour of the fluid to be continuous across scales), we assume that the fields $\boldsymbol{W}_{\ell}^{0}$ are filtered versions of a single random field $\boldsymbol{W}^{0}$. 

We approximate the evolution equation one step further by using the advection velocity field $\boldsymbol{\tilde{v}}_{\bigell}$ computed from the synthetic state variable $\boldsymbol{\tilde{W}}_{\bigell} \equiv \int_{s\bigell}{\rm d}\ln s\,\boldsymbol{\tilde{W}}_{s}$ in equation \eqref{eq:Wtildel}, and similarly for $\boldsymbol{\boldsymbol{S}}_{\bigell}$, for which we use $\boldsymbol{\boldsymbol{\tilde{S}}}_{\bigell}$.

Finally, the line of reasoning detailed above ultimately leads to the following approximate evolution equation for the dynamics at a given scale $\ell$
\begin{equation}
\label{eq:Wtildel-vtilde}
\partial_{t}\boldsymbol{\tilde{W}}_{\ell}\simeq
-\boldsymbol{\tilde{v}}_{\bigell}\boldsymbol{.\nabla\tilde{W}}_{\ell}
+\boldsymbol{\tilde{S}}_{\bigell}\boldsymbol{.\tilde{W}}_{\ell}
+\boldsymbol{D}[\boldsymbol{\tilde{W}}_{\ell}].
\end{equation}
Interestingly, it is almost in the initial generic form \eqref{eq:generic-form}, except that the advection velocity and deformation matrix now depend on all scales above $\ell$.\\

In the following, we will introduce yet more approximations, now tailored to \textit{solve} approximately the approximate evolution equation \eqref{eq:Wtildel-vtilde} we just derived, i.e. we will present our turbulence synthesis procedure. At this stage, the reader may legitimately be worried that such an accumulation of (even small) losses of information ought to be too drastic, and can eventually only move us far away from the correct evolution. If so, we invite the reader to recall figure~\ref{fig:teaser} (or to have a glimpse at the more detailed figures \ref{fig:vorticity-maps} and \ref{fig:maps-6}) were we showed a non-trivial case were the synthesis and the full simulation do match very well. Therefore, to some extent the whole procedure does keep enough relevant information to be of practical use, i.e. we have meaningful examples where it seems to extract the essence of the dynamics.

\subsection{Turbulence synthesis}
\label{sec:synthesis}
Our last equation \eqref{eq:Wtildel-vtilde} is in closed form for all scales above $\ell$: the evolution for $\boldsymbol{\tilde{W}}$ at scale $\ell$ can be solved once all scales $s$ \emph{above} $\ell$ are known. We can hence sweep all scales from the largest down to the smallest and compute $\boldsymbol{\tilde{W}}_{\ell}$ at all scales $\ell$ (we will discuss how we deal with the largest scale in the numerical implementation below, see section \ref{sec:numerical}). 
The resulting synthesised field can finally be reconstructed as $\boldsymbol{\tilde{W}}=\int{\rm d}\ln\ell \,\boldsymbol{\tilde{W}}_{\ell}$.

Variants of our model will now be defined by the way we approximate the evolution equation \eqref{eq:Wtildel-vtilde} (detailed in section \ref{subsec:Time-evolution}) and the way we estimate the coherence time $\tau_{\ell}$ (developed in section \ref{subsec:coherence-time}).

\subsubsection{Time evolution approximations}
\label{subsec:Time-evolution}

Here, we focus on our approximations to solve the evolution equation \eqref{eq:Wtildel-vtilde}.
We assume that the fields $\boldsymbol{\tilde{v}}_{\bigell}$ and $\boldsymbol{\tilde{S}}_{\bigell}$ evolve slowly compared to the fields at scale $\ell$, as they are based on larger scales.  
We therefore assume they remain constant during the time interval $\tau_{\ell}$. For instance, we assume that the fluid parcels move ballistically with the velocity $\boldsymbol{\tilde{v}}_{\bigell}$ during a coherence time $\tau_{\ell}$ to integrate \eqref{eq:Wtildel-vtilde} in time.

We first focus on a case where $\boldsymbol{S}=\boldsymbol{D}=\boldsymbol{0}$ and $\boldsymbol{v}\neq\boldsymbol{0}$. In this case, we only need to randomly draw the initial vorticity field $\boldsymbol{W}_{\ell}^{0}$ at time $t_{\ell}^{0}$ and remap it at positions $\boldsymbol{x}^{0}=\boldsymbol{x}-\tau_{\ell}\boldsymbol{\tilde{v}}_{\bigell}$ where the fluid parcel used to be:
\begin{equation}
\boldsymbol{\tilde{W}}_{\ell}^{\boldsymbol{v}\neq\boldsymbol{0}}(\boldsymbol{x})=\boldsymbol{W}_{\ell}^{0}(\boldsymbol{x}-\tau_{\ell}\boldsymbol{\tilde{v}}_{\bigell}).
\end{equation}
Possible shell crossing will be discussed in the next subsection \ref{subsec:coherence-time}.

We now focus on the case $\boldsymbol{v}=\boldsymbol{D}=\boldsymbol{0}$ ($\boldsymbol{S}\neq\boldsymbol{0}$). Since $\boldsymbol{\tilde{S}}_{\bigell}$ is constant during $\tau_\ell$, the solution of $\partial_{t}\boldsymbol{\tilde{W}}_{\ell}=\boldsymbol{\tilde{S}}_{\bigell}.\boldsymbol{\tilde{W}}_{\ell}$ is obtained as a matrix exponential :
\begin{equation}
\boldsymbol{\tilde{W}}_{\ell}^{\boldsymbol{S}\neq\boldsymbol{0}}(\boldsymbol{x})=\exp(\tau_{\ell}\boldsymbol{\tilde{S}}_{\bigell}).\boldsymbol{W}_{\ell}^{0}(\boldsymbol{x}).\label{eq:only-stretch}
\end{equation}

For the general case, we combine the two previous steps as 
\begin{equation}
\boldsymbol{\tilde{W}}_{\ell}^{\boldsymbol{D}=\boldsymbol{0}}(\boldsymbol{x})=\exp(\tau_{\ell}\boldsymbol{\tilde{S}}_{\bigell}).\boldsymbol{W}_{\ell}^{0}(\boldsymbol{x}-\tau_{\ell}\boldsymbol{\tilde{v}}_{\bigell}),\label{eq:advection-deformation}
\end{equation}
and we finally apply the diffusion step as
\begin{equation}
\boldsymbol{\tilde{W}}_{\ell}\simeq\boldsymbol{\tilde{W}}_{\ell}^{\boldsymbol{D}=\boldsymbol{0}}+\tau_{\ell}\boldsymbol{D[}\boldsymbol{\tilde{W}}_{\ell}^{\boldsymbol{D}=\boldsymbol{0}}\boldsymbol{]}.\label{eq:diffusion}
\end{equation}

We take those three steps in a particular order: advection, deformation, then diffusion. We feel it is justified by the need to use the quantities $\boldsymbol{v}$ and $\boldsymbol{S}$ at a time as advanced as possible (i.e. as close as possible to the time at which we compute the synthesised flow). Also the computation of the gradients in the deformation matrix requires a synchronous evaluation, so deformation might be better evaluated after advection.
We finish by the diffusion step because it somewhat smooths the fronts generated at the convergence points of advection trajectories and where the deformation introduces stretching (a property sought for in adhesion models, for example, see \citealp{Gurbatov1989}).  This also allows to alleviate the small scales generated outside the current filter by the non-linear action of advection and deformation (see discussion in section \ref{subsec:small-terms}, also).
However, in the limit where $\tau_\ell$ is very small, time evolution is linear and the order does not matter. 

\subsubsection{Coherence time estimates}

\label{subsec:coherence-time}
So far, we have defined the coherence time in this work as the time during which the evolution of the fluid can be considered deterministic at a given length scale. We will now examine three ways to estimate this time.

1. It is at first tempting to identify it to the correlation time scale. 
For example, in the direct enstrophy cascade of 2D turbulence, below the injection scale, the correlation time scale is constant, equal to the injection time scale. It would then make sense to adopt $\tau_\ell=\rm constant$ at all scales. 

2. The notion of coherence also appeals to the sensitivity to initial conditions. The rate at which fluid trajectories deviate from each other is linked to the symmetric part of the velocity gradient matrix, which characterises the evolution of distances between nearby fluid parcels. We define the stretching time as the inverse of its largest eigenvalue which in incompressible fluids writes as \citep{Haller2021}:
\begin{equation}
\label{eq:stretch}
    \tau^{-2}_{\rm stretch}(\boldsymbol{u})=
    |\partial_x u_{x}\partial_y u_{y}
    -\frac14(\partial_x u_{y}+\partial_y u_{x})|.
\end{equation}
This eigenvalue can also be seen as the largest instantaneous Lyapunov exponent for fluid trajectories. 
The above expression for incompressible hydrodynamics  (in which $\partial_x u_{x}=-\partial_y u_{y}$) can be recast into 
\begin{equation}
\label{eq:strain}
    4\tau^{-2}_{\rm stretch}(\boldsymbol{u})=
    (\partial_x u_{x}-\partial_y u_{y})^2
    +(\partial_x u_{y}+\partial_y u_{x})^2=\tau^{-2}_{\rm strain},
\end{equation}
which expresses the stretch rate as a sum of squares of the normal and shear strains. That time scale is hence also naturally related to the deformation of fluid elements. Indeed, in some places the fluid might be rotating in a vorticity clump and its evolution at scales below the size of that vortex could be considered deterministic. On the contrary, fluid parcels in between vortices experience strong deformation through shearing motions. Adopting $\tau_\ell=\tau_{\rm stretch}(\boldsymbol{u}_\ell)$ would also ensure that fluid elements retain their integrity (i.e. they are not too deformed and it still makes sense to follow the same fluid particle).

3. Since we resort to a ballistic approximation for the advection of the fluid, our modeled fluid trajectories may cross each other after some finite time. This shell-crossing time can be defined as the inverse of the largest eigenvalue of the velocity gradient (by contrast to its symmetric part above):
\begin{equation}
\label{eq:shell-crossing}
    \tau^{-2}_{\rm shell}(\boldsymbol{u})=
    |\partial_x u_{x}\partial_y u_{y} -
    \partial_x u_{y}\partial_y u_{x}|
    .
\end{equation}
Integration of ballistic trajectories for more than this time scale does not make sense.
Note that this constraint is more related to our ballistic approximation than to the stochastic behaviour of the fluid, and may be relaxed once we resort to a higher order approximation. However, in the case of 2D incompressible hydrodynamics, this time scale is related to the physically motivated Okubo-Weiss \citep{Okubo1970,Weiss1991,Haller2021,Shivamoggi2024} criterion $\tau^{-2}_{\rm shell}=|\tau^{-2}_{\rm stretch}-w^2|$. That last expression can also be used to show that requesting $\tau_{\ell}=\tau_{\rm shell}$ is slightly less stringent than the stretch criterion ($\tau_{\rm stretch}\le\tau_{\rm shell}$). 

To conclude, because it is not yet clear which of these approaches is best suited, we have tested four choices for the coherence time (from less to more stringent ($\tau_{\rm strain}<\tau_{\rm stretch}\le\tau_{\rm shell}$ ):
\begin{itemize}
    \item Constant $\tau_\ell=t$
    \item Shell-crossing based $\tau_\ell^{-2}=t^{-2}+\tau_{\rm shell}(\boldsymbol{u}_{\bigell})^{-2}$
    \item Stretch based $\tau_\ell^{-2}=t^{-2}+\tau_{\rm stretch}(\boldsymbol{u}_{\bigell})^{-2}$.
    \item Strain based $\tau_\ell^{-2}=t^{-2}+\tau_{\rm strain}(\boldsymbol{u}_{\bigell})^{-2}$.
\end{itemize}
The minus-two powers in the above expressions take soft minima between the various time scales involved and the integration time $t$. Indeed, when we compare our method to actual simulations, we compute the evolution of the fluid over a finite time $t$ from random initial conditions. In this case, it is not necessary to integrate the evolution of the synthesised flow over times longer than $t$. 

As a final note, most of these definitions imply that the coherence time $\tau_\ell(\boldsymbol{x},t)$ depends on both scale, position in space and total evolution time $t$ as long as this one is not too long compared to local times of the flow. The initial conditions $\boldsymbol{W}_{\ell}^{0}$ should hence not be understood as starting conditions at a synchronous time. Rather, they are random parameters which reflect the sensitivity to initial conditions. 

\subsection{Numerical implementation}
\label{sec:numerical}

\subsubsection{Discrete filters}

A numerical implementation of the above method requires a space domain with a finite extent and a finite space resolution. As a result, the Fourier domain spans a finite number of wavenumbers, and only a finite number of scales is necessary to recover a given field. We choose to consider a cubic (square in 2D) domain of size $L=2\pi$ gridded with $N$ pixels and we adopt periodic boundary conditions. The available wavevectors therefore span a regular grid of unit $k_{J}=2\pi/L=1$. We will restrict ourselves to a discrete set of $J+1$ filters $\left\{ \varphi_{j}\right\} _{j=0,..,J}$ logarithmically distributed in scales in such a way that 

\begin{equation}
\hat{\varphi}_{j+1}(k)=\hat{\varphi}_{j}(k/\lambda),
\end{equation}
with $\lambda\in]0,1[$ the ratio between the typical scales of two consecutive filters. This parameter controls the resolution in scales. In the present application, we mostly use $\lambda=1/2$ but we also tested a finer set of scales with $\lambda=1/\sqrt{2}$ (hence with twice as many filters), without noticing much change.

We request each filter to be selective in logarithmic scales, in order to be able to use the scale separation arguments at the base of our approximations. The filters $\hat{\varphi}_{j}$ hence select concentric coronas in Fourier space and are centred around wavenumbers $k_{j}=k_{J}\lambda^{j-J}$ with a relative width of $1/\lambda$. In other words, the typical scale associated with filter $j$ is 
\begin{equation}
    \ell_{j}=L\lambda^{J-j}. 
\end{equation}
We point out that these physical scales go in increasing order from the smallest to the largest as $j$ increases.

The discretised version of our normalisation condition \eqref{eq:normalisation} will simply be to enforce $\sum_{j=0}^{J}\hat{\varphi}_{j}=1$ for any $\vec{k}$ contained in the computational domain. Indeed, then the reconstruction formula works as 
\begin{equation}
\sum_{j=0}^{J}\hat{f}_{j}=\hat{f}.\sum_{j=0}^{J}\hat{\varphi}_{j}=\hat{f},
\end{equation}
 from which we deduce $f=\sum_{j=0}^{J}f_{j}$.
We present two example filters we used in the present application in the appendix \ref{sec:filters}.

\subsubsection{Algorithm}

The initial Gaussian field $\boldsymbol{W}^{0}$ is first drawn in real space as a Gaussian white noise. We then convolve it with the desired spectrum by multiplying its Fourier coefficients. 

We present below the steps of the algorithm we use to generate turbulent fields $\boldsymbol{\tilde{W}}_{j}$ at all scales $j$ from the initial Gaussian field $\boldsymbol{W}^{0}$. We write $\boldsymbol{\tilde{W}}_{j}=\boldsymbol{\tilde{W}}*\hat{\varphi_{j}}$ and $\boldsymbol{W}_{j}^{0}=\boldsymbol{W}^{0}*\hat{\varphi_{j}}$.

To initiate the first $j=J$ step of our iterative algorithm, at the largest scale, we simply assume $\boldsymbol{\tilde{W}_{J}}=\boldsymbol{W}_{j=J}^{0}$. We then parse the successive scales downward, and we successively apply the following operations at each scale:

\begin{enumerate}

\item At stage $j<J$, the fields $\boldsymbol{\tilde{W}}_{i}$
are known for all $i>j$.

\item We write $\boldsymbol{\tilde{W}}_{\geqslant j}=\sum_{i>j}\boldsymbol{\tilde{W}}_{i}+\boldsymbol{W}_{j}^{0}.$
We compute the advection velocity $\boldsymbol{v}_{\geqslant j}=\boldsymbol{v}[\boldsymbol{\tilde{W}}_{\geqslant j}]$
and similarly $\boldsymbol{S}_{\geqslant j}=\boldsymbol{S}[\boldsymbol{\tilde{W}}_{\geqslant j}]$.

\item We estimate $\tau_{j}=\tau_{\ell_{j}}$ (for which we might need $\boldsymbol{v}_{\geqslant j}$
if we use the local stretch or shell crossing times, for example).

\item We now replace these values in the construction formulas \eqref{eq:advection-deformation}
and \eqref{eq:diffusion}. We compute the time-evolved field at scale
$\ell_{j}$ (without diffusion) as 
\begin{equation}
\boldsymbol{\tilde{W}}_{j}^{\boldsymbol{D}=\boldsymbol{0}}(\boldsymbol{x})=\exp(\tau_{j}\boldsymbol{\tilde{S}}_{\geqslant j}(\boldsymbol{x})).\boldsymbol{W}_{j}^{0}(\boldsymbol{x}-\tau_{j}\boldsymbol{v}_{\geqslant j})
\end{equation}
where we use a first order linear interpolation scheme to compute
$\boldsymbol{W}_{j}^{0}$ at positions in between grid points. 

\item We finally apply diffusion to the resulting field to get $\boldsymbol{\tilde{W}}_{j}=\boldsymbol{\tilde{W}}_{j}^{\boldsymbol{D}=\boldsymbol{0}}+\tau_{j}\boldsymbol{D}[\boldsymbol{\tilde{W}}_{j}^{\boldsymbol{D}=\boldsymbol{0}}].$
We now have $\boldsymbol{\tilde{W}}_{i}$ known for all $i>j-1$ and
we can proceed further back to operation 1 down to the next scale $j-1$.

\end{enumerate}

The loop stops when we hit the smallest scale labelled by $j=0$.
The final field then results from $\boldsymbol{\tilde{W}}=\sum_{j=0}^{J}\boldsymbol{\tilde{W}}_{j}$.

For the interpolation at operation 4, we use  \texttt{RegularGridInterpolator} from the \texttt{scipy.interplolate} \texttt{Python} package. One of the caveats of this
interpolation remapping is that it is not divergence and mean preserving. One may need to reproject the divergence free fields and adjust preserved quantities after the interpolation process. In our 2D hydrodynamical application in section \ref{sec:validation}, we remove its average from the vorticity field to recover the zero mean. The same holds for the deformation step.  

The matrix exponential at operation 4 makes use of the Pad\'e approximant implementation from the EXPOKIT package \citep{EXPOKIT}, for which we implemented a vectorised wrapper in
order to gain efficiency (essentially to call it from \texttt{Python} in
a more efficient way for an array of vectors).

Diffusion at operation 5 is computed in Fourier assuming it proceeds over a homogenised time $\bar{\tau}_{j}$ averaged over the computational domain.
In this simplified case, the effect of incompressible viscous diffusion $\partial_{t}\boldsymbol{w}=\boldsymbol{D[w]}=\nu\Delta\boldsymbol{w}$ during time $\bar{\tau}_{j}$ is simply to multiply by $\exp(-\nu\,k^{2}\bar{\tau}_{j})$ the Fourier coefficients of the initial field $\boldsymbol{w}$.
\tcg{Note that because $\tau_j$ is not necessarily uniform over the computational domain, we resort to using a volume averaged value for $\tau_j$. Otherwise, there is to our knowledge no convenient way to integrate the viscous diffusion term.}

\subsubsection{CPU cost scaling}

If we neglect the cost of the matrix exponentiation and the remapping (proportional to the number of pixels), most of the computational cost will amount to Fourier transforms.
  At each scale index $j>0$ we need to Fourier transform back and
forth because some operations are more efficiently performed in Fourier
space (computing the advection velocity, the deformation matrix, or applying the diffusion process) while others are better done in real space (interpolation remap and deformation). 
To be more precise, the 2D synthesis uses 10 calls to the fast Fourier transform (FFT) for each scale level: 2 for the advection velocity, 4 for the deformation matrix, 2 around the remap step and 2 around the diffusion step.
A 3D synthesis would require 3 for the advection velocity, 9 for the deformation matrix, and 6 for each of the remap and diffusion step: a total of 24 calls to a scalar FFT.
In general, the method hence requires $10J$ (2D) or $24J$ (3D) Fourier transforms to generate one field. 
Since the CPU cost of a Fourier transform in $d$ dimensions is on the order of $N^d\log_2 N^d$ operations and $J$ scales like $\log_2 N$, the cost of one field generation scales as 
\begin{equation}
    N_{\rm 2D syn}=20 N^2(\log_2 N)^2
\end{equation} operations in 2D and 
\begin{equation}
    N_{\rm 3D syn}=72 N^3(\log_2 N)^2
\end{equation} in 3D.

The cost of one time step of a pseudo-spectral code is proportional to the order of the time integrator: a Runge-Kutta order 4 such as we use in section \ref{sec:reference-simulations} requires 4 evaluations of the time derivative of the quantities at play. 
Each of these evaluation costs 5 FFTs in 2D (2 for the velocity, 2 for the diagonal components of the vorticity gradient, 1 to get back to Fourier space), and 3+3+9+3=18 in 3D where the whole velocity gradient terms are also needed. 
At a Courant number of 1, the Courant-Friedriech-Lewy condition constrains the time-step to be less than $t_{\rm turnover}/N$: one requires $N$ time steps to reach one turnover time. The cost of a 2D simulation is hence on the order of 
\begin{equation}
    N_{\rm 2D sim}=40 N^{3}\log_2 N\,.t/t_{{\rm turnover}}
\end{equation} operations for our Runge-Kutta order 4 implementation, and  
\begin{equation}
    N_{\rm 3D sim}=216 N^{4}\log_2 N\,.t/t_{{\rm turnover}}
\end{equation} in 3D.

\begin{figure}
    \centering
    \includegraphics[width=\columnwidth]{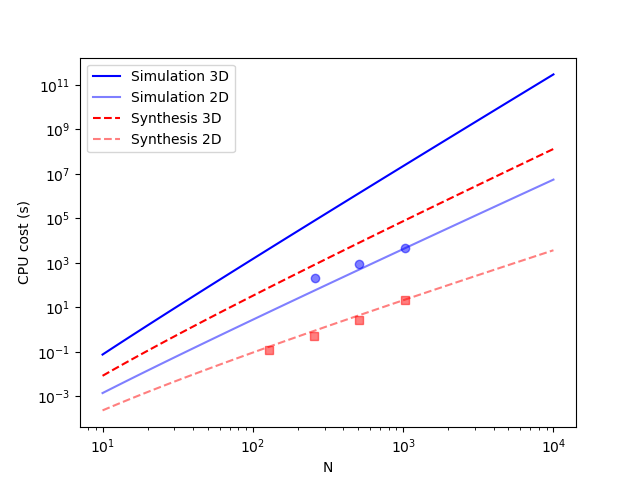}
    \caption{CPU cost scaling estimates for 2D (light colors) and 3D (dark colors) hydrodynamics simulations (solid, blue) and MuScaTS syntheses (dashed, red). The scalings are computed according to text, calibrated on a 2D simulation (pseudo-spectral code implemented in Fortran) at $N=1024$ for one turnover time (right most blue circle). The total CPU cost for simulations at $N=256$ and $N=512$ (the other two blue circles) are also indicated. Red squares stand for the CPU cost of a few 2D MuScaTS syntheses.}
    \label{fig:CPU-cost}
\end{figure}

  For a simulation over the length of one turnover time, the CPU time gain factor is hence $2N/\log_2(N)$ in 2D and $3N/\log_2(N)$ in 3D. Both 2D and 3D scalings for simulations and MuScaTS syntheses are displayed on figure \ref{fig:CPU-cost}. 
  The scalings are all calibrated on a 2D  $N=1024$ simulation run (rightmost blue circle) using our pseudo-spectral code (see section \ref{sec:reference-simulations}) compiled with Fortran on the \texttt{totoro} server at the mesoscale centre mesoPSL. For reference, the CPU hardware on this server is Intel(R) Xeon(R) Gold 6138 CPU, 2.00GHz.
  Other circles indicate the scaling of simulations at $N=256$ and $N=512$: they scale slightly less well than the theory predicts, because we ran our code in parallel on 32 cores, so the communication overhead in the FFT at smaller computational domains (smaller $N$) deteriorates the scaling. The light red squares show the CPU cost on the same hardware for a python 2D implementation of our MuScaTS synthesis.

 As a final note, we point out that the current implementation could be further optimised with respect to the multiscale treatment. For a proper choice of filters, we could indeed afford using coarser grids at larger scales than for the smaller scales. This would considerably accelerate the computation at largest scales and lower the number of effective number of scales which enter the CPU cost scaling.

\section{Comparison to other methods}

We investigate how our MuScaTS technique compares to selected existing works. 

\subsection{Comparison to \citet{Rosales2006}}

For 3D hydrodynamics, if we neglect the deformation and the diffusion
terms, our method is extremely similar to the minimal multiscale Lagrangian
map (MMLM) of \citet{Rosales2006} provided we use $\tau_{\ell}$
as in their equation (9): $\tau_{\ell}^{{\rm MMLM}}=\ell/u_{{\rm r.m.s.}\ell}$
where $u_{{\rm r.m.s.}\ell}$ is the prescribed r.m.s. velocity in
the initial field at scale $\ell$. The only remaining difference is:
we would advect the vorticity field and build the velocity from a
Biot-Savart formula, while their Lagrangian map is directly applied to the
velocity. They
have to reproject their velocity field on $\boldsymbol{\nabla.u}=0$ while we get
divergence free fields by construction. We hence expect our method
should be more realistic, as both vorticity stretching effects and diffusion are retained while in effect, they neglect pressure terms and diffusion.

In a later paper, the same authors \citet{2008PhRvE..78a6313R} noticed
that they could improve their method by cutting single Lagrangian
steps into smaller ones as scales become finer. They considerably
improved the realism of the anomalous exponents using this approach,
which they call multiscale turnover Lagrangian map (MTLM for short).
We note that we could also incorporate this multi-stepping into our
MuScaTS method, although it would increase its cost. Our method could perhaps also benefit from
the fast algorithms designed by \citet{2016PhRvE..94e3109M} to improve
the efficiency of the MTLM method.

Finally, \citet{Subedi2014} designed an incompressible MHD
version of the MMLM method. MuScaTS without diffusion or deformation
applied to the choice of variable $\boldsymbol{W}=(\boldsymbol{\nabla}\times\boldsymbol{u},\boldsymbol{B})$
and using $\tau_{\ell}=\tau_{\ell}^{{\rm MMLM}}$ should amount to
the same model except we advance the vorticity $\boldsymbol{\nabla}\times\boldsymbol{u}$
rather than the velocity, which  directly generates divergence
free fields (we would however still have to clean the divergence of the
magnetic field). The extension of \citet{Subedi2014} by \cite{Luebke2024} proceeds with advection from an already non-Gaussian field and produced convincing coherent structures. Our own implementation will advect along an already deformed field which contains physically motivated non-Gaussianities: we hence expect that the MHD version of MuScaTS should perform equally well if not better, but this remains to be demonstrated.

\subsection{Comparison to \citet{Chevillard2011} (CRV)}

The link between MuScaTS and CRV is slightly less direct and we need to rework the equations quite a bit to highlight the similarities and differences.

First, let us expose part of our model in a way that will ease the comparison.
The work of CRV is in the context of 3D Navier-Stokes incompressible equations, so in our framework we take $\boldsymbol{W}=(\boldsymbol{w})$ where $\boldsymbol{w}$ is the vorticity.
We neglect diffusion, or in other words we focus on the inertial range. If we
now neglect advection rather than deformation, as in CRV, then only the exponential
term remains in equation \eqref{eq:only-stretch} which we rewrite
more precisely here as
\begin{equation}
\boldsymbol{\tilde{w}}_{\ell}^{\boldsymbol{S}\neq0}(\boldsymbol{x})=\exp(\tau_{\ell}\boldsymbol{S}[\boldsymbol{\tilde{w}}_{\bigell}\boldsymbol{]}).\boldsymbol{w}_{\ell}^{0}(\boldsymbol{x}),
\label{eq:only-stretch-1}
\end{equation}
to make clear the intricate dependence of $\boldsymbol{S}$ on the
larger scales. If to parallel the work of CRV we now approximate the argument of $\boldsymbol{S}$
by its value ab initio, we get 
\begin{equation}
\boldsymbol{\tilde{w}}_{\ell}(\boldsymbol{x})=\exp(\tau_{\ell}\boldsymbol{S}[\boldsymbol{w}_{\bigell}^{0}\boldsymbol{]}).\boldsymbol{w}_{\ell}^{0}(\boldsymbol{x})
\end{equation}
or now using equation \eqref{eq:bigell}, we obtain
\begin{equation}
\label{eq:no-adv-local}
\boldsymbol{\tilde{w}}_{\ell}(\boldsymbol{x})=\exp \left( \int_{s\bigell}{\rm d}\ln s \ \! \tau_{\ell} \ \! \boldsymbol{S}[\boldsymbol{w}_{s}^{0}\boldsymbol{]} \right).\boldsymbol{w}_{\ell}^{0}(\boldsymbol{x}),
\end{equation}
and, integrating over all scales, we finally get
\begin{equation}
\boldsymbol{\tilde{w}}^\mathrm{MuScaTS}=\int_{-\infty}^{+\infty}{\rm d}\ln \ell \exp \left( \int_{s\bigell}{\rm d}\ln s \ \! \tau_{\ell} \ \! \boldsymbol{S}[\boldsymbol{w}^{0}_{s}\boldsymbol{]} \right).\boldsymbol{w}_{\ell}^{0}.
\label{eq:MuScaTS-CRV}
\end{equation}
This formula will be our comparison point to CRV's work below.

Then, let us reformulate part of CRV's model. In their work, they construct a non-Gaussian random velocity field as the result of the vortex stretching of some initial Gaussian vorticity field $\boldsymbol{w}^0$. Concretely, they consider for $\boldsymbol{w}^0$ a Gaussian white noise vector (that we will here denote $\boldsymbol{\eta}$) and insert into a Biot-Savart-type formula a proxy for the vorticity field of the form
\begin{equation}
\boldsymbol{w}^{{\rm CRV}}=\exp(\tau_{\rm CRV} \boldsymbol{S}^{\rm CRV}[\boldsymbol{w}^0]).\boldsymbol{w}^{0},
\label{eq:CRV0}
\end{equation}
where $\tau_{\rm CRV}$ is a parameter which controls the amount of stretching, as it relates\footnote{ \tcg{The focus in this section is on the scale dependency of parameters (power-law variation in $\ell$) so throughout this section we omit numerical factors, as in \eqref{eq:S_generalExpression}, but it is worth mentioning that strictly speaking, $\tau_{\rm CRV}$ does not have the dimension of a time. It is a parameter controlling the intermittency (non-Gaussianity) of the resulting velocity field which appears in CRV's construction from an equation governing the dynamics of stretching, so we choose this $\tau$ notation to help the reader get some intuition, notably since the longer stretching occurs, the larger this parameter is. For more details see CRV's work \citep{Chevillard2011, Chevillard15_HDR}.} } to the correlation time-scale of the velocity gradients, and $\boldsymbol{S}^{\rm CRV}[\boldsymbol{w}^0]$ is a matrix field encoding the stretching of $\boldsymbol{w}^0$ during a typical time $\tau_{\rm CRV}$.
Now, as a general rule, given a vorticity field $\boldsymbol{w}$ the stretch matrix field reads\footnote{For simplicity and since we focus on the inertial range in this discussion, we omitted the numerical prefactor, the regularisation and the principal value present in CRV.}
\begin{equation}
\boldsymbol{S}[\boldsymbol{w}](\boldsymbol{x})\propto\int{\rm d}^{3}\boldsymbol{y}
\frac{\boldsymbol{r}\otimes(\boldsymbol{r}\times\boldsymbol{w}(\boldsymbol{y}))
+(\boldsymbol{r}\times\boldsymbol{w}(\boldsymbol{y}))\otimes\boldsymbol{r}}
{r^{5}},
\label{eq:S_generalExpression}
\end{equation}
where $\boldsymbol{r} \equiv \boldsymbol{x-y}$ and $\otimes$ denotes the tensor product. In addition, an important feature which is universally observed in both experiments and numerical simulations of turbulence, is that turbulent fields are multi-fractal and have long-range correlations. This formally translates into $\boldsymbol{S}[\boldsymbol{w}]$ being log-correlated in space, and a key point in CRV's work which greatly increases the realism of their model, is to implement this feature into their formulae through a simple, effective rule\footnote{In follow-ups of their CRV work, such as \citet{Chevillard15_HDR, PereiraEtAl18}, they do support this with very rigorous but rather lengthy derivations, and we here stick to their simple reasoning yielding the correct result }, namely by just changing the exponent in the power-law $r^5$ of the denominator of the general form \eqref{eq:S_generalExpression}.
Specifically, by choosing $\boldsymbol{\eta}$ for their $\boldsymbol{w}^0$ they adopt the rule $r^5 \rightarrow r^{7/2}$, i.e. they consider (cf. CRV, equation 12)
\begin{equation}
\boldsymbol{S}^{{\rm CRV}}[\boldsymbol{\eta}](\boldsymbol{x})\propto\int{\rm d}^{3}\boldsymbol{y}
\frac{\boldsymbol{r}\otimes(\boldsymbol{r}\times\boldsymbol{\eta}(\boldsymbol{y}))
+(\boldsymbol{r}\times\boldsymbol{\eta}(\boldsymbol{y}))\otimes\boldsymbol{r}
}
{r^{7/2}},
\label{eq:CRV_7over2}
\end{equation}
and this simple tweak makes their stretch field log-correlated as it should.

Hence, in CRV's work, log-correlation appears through a cunning but somewhat artificial change of exponent, while the stretching time-scale $\tau_{\rm CRV}$ remains a mere parameter: in particular it is scale-independent. Inspired by their line of thinking, we now suggest an alternative approach in which we retain the physical scaling for the stretch, but we now introduce a scale-dependency of the stretching time-scale to achieve log-correlation. Concretely, starting from expressions \eqref{eq:CRV0} and \eqref{eq:S_generalExpression}, and imposing log-correlation with a similar scaling argument as CRV's, we now build expression \eqref{eq:CRV} below, which is reminiscent of the scale-decomposition \eqref{eq:MuScaTS-CRV} of our model and may thus be interpreted in terms of a scale-dependent stretching time-scale.

To improve on the physical interpretation of CRV, instead of choosing as initial condition $\boldsymbol{w}^0$ the simple Gaussian white noise $\boldsymbol{\eta}$ we also adopt a slightly more sophisticated Gaussian field $\boldsymbol{w}^{0}$ with the \cite{K41} power spectrum of the vorticity (i.e. with a spectrum scaling as $E_{w^{2}}(k)\sim k^{1/3}$). But doing so first raises the question: how is CRV's effective rule then modified? To answer this, we suggest the following rough scaling argument.
In Fourier space, for a given corona of radius $k$ and infinitesimal thickness ${\rm d}k$, by definition of the spectrum $E_{w^{2}}(k){\rm d}k=\hat{w}^2k^2{\rm d}k$, so our choice of Kolmogorov scaling implies $\hat{w}\sim k^{-5/6}$. 
Then, since the Fourier transform of a power law $r^{-\alpha}$ is $k^{\alpha-d}$ where $d=3$ is the dimension of space, our scaling in Fourier space corresponds to a scaling $w\sim r^{5/6-3}$ in position space. Hence, compared to CRV's case, we have an additional power $r^{5/6-3}$ in the numerator of the ratio in \eqref{eq:CRV_7over2}, so log-correlation requires that we adjust the exponent in the denominator to compensate it, i.e. we consider
\begin{equation}
\boldsymbol{S}^{{\rm CRV}}[\boldsymbol{w}^{0}](\boldsymbol{x})\propto\int{\rm d}^{3}\boldsymbol{y}
\frac{\boldsymbol{r}\otimes(\boldsymbol{r}\times\boldsymbol{w^{0}}(\boldsymbol{y}))
+(\boldsymbol{r}\times\boldsymbol{w^0}(\boldsymbol{y}))\otimes\boldsymbol{r}
}
{r^{4/3}},
\label{eq:S_4over3}
\end{equation}
where the denominator should be understood as $r^{7/2+(5/6-3)} = r^{4/3}$.
This is our modification to CRV's effective rule.

Now, to draw a parallel with the MuScaTS construction, let us build this stretch matrix $\boldsymbol{S}^{{\rm CRV}}[\boldsymbol{w}^{0}]$ from a continuous series of narrow filters, by decomposing it into various scales with \eqref{eq:reconstruction} and using the linearity property \eqref{eq:linearity}. We thus have
\begin{equation}
\boldsymbol{S}^{{\rm CRV}}[\boldsymbol{w}^{0}]=\int_{-\infty}^{+\infty}{\rm d}\ln\ell \ \boldsymbol{S}^{{\rm CRV}}[\boldsymbol{w}_{\ell}^{0}],
\label{eq:S_CRV_scaleDecomposition}
\end{equation}
but in fact, to be complete, since in CRV's model what drives the dynamics is the full argument of the matrix exponential in \eqref{eq:CRV0}, we also need to take into account the $\tau_{\rm CRV}$ parameter. Here instead of this mere parameter, we introduce some scale-dependent $\tau_\ell$ in the scale-decomposition \eqref{eq:S_CRV_scaleDecomposition} such that
\begin{equation}
\tau_{\rm CRV} \boldsymbol{S}^{{\rm CRV}}[\boldsymbol{w}^{0}] = \int_{-\infty}^{+\infty}{\rm d}\ln\ell\, \tau_\ell \ \boldsymbol{S}[\boldsymbol{w}_{\ell}^{0}],
\label{eq:tau_S_CRV}
\end{equation}
where $\boldsymbol{S}$ is the real, physical stretch \eqref{eq:S_generalExpression}, i.e. with a $r^5$ denominator. Then, to reconcile the $r^{4/3}$ requirement for log-correlation of \eqref{eq:S_4over3} with the real physical $r^5$ dependency of the stretching, we argue (pursuing the hand-wavy reasoning consisting in keeping track and matching length scales) that a power-law dependency
\begin{equation}
    \tau_\ell \propto \ell^{5 - 4/3} = \ell^{11/3},
    \label{eq:tau_ell_CRV}
\end{equation}
is adequate.
Finally, using in \eqref{eq:CRV0} expression \eqref{eq:tau_S_CRV} and the scale decomposition \eqref{eq:reconstruction} for $\boldsymbol{w}^{0}$, we rewrite the CRV random field in the following form, which is close to our \eqref{eq:MuScaTS-CRV},
\begin{equation}
\boldsymbol{w}^{{\rm CRV}}=\int_{-\infty}^{+\infty}{\rm d}\ln \ell \exp \left( \int_{-\infty}^{+\infty}{\rm d\ln s\,} \ \! \tau_{s} \ \! \boldsymbol{S}[\boldsymbol{w}_{s}^{0}\boldsymbol{]} \right).\boldsymbol{w}_{\ell}^{0},
\label{eq:CRV}
\end{equation}
with $\tau_{\ell}$ given by \eqref{eq:tau_ell_CRV} when the vorticity has a \cite{K41} power-law spectrum. It is in this sense that we mean that the change of power-law to enforce log-correlation in CRV could be interpreted as an integration time varying with scale, with a typical time scale $\tau_{\ell}\propto\ell^{11/3}$.

This unusual perspective on their theory allows to explicit more
of the original physical grounds behind their construction. 
Thanks to the freedom introduced by the scale dependent stretch parameter, we can both retain the physical expression for the stretch (with $1/r^5$ scaling) and assimilate the generating Gaussian field as the initial vorticity (using Kolmogorov scaling instead of a white noise).

Now, comparing the above reformulations of both formalisms, we reveal three noteworthy differences.
First, as can be seen in \eqref{eq:no-adv-local}, we use a stretch field in the larger scales which has already been advanced, while CRV uses the initial stretch field at all scales.
Second, \eqref{eq:MuScaTS-CRV} confronted to \eqref{eq:CRV} shows that in our formulation the effective time scale $\tau_\ell$ applies to all scales above while it is local in scales in CRV. 
Third, the integration bounds differ: we integrate only the scales above $\ell$ while CRV uses the whole range of scales. In other words, our method suppresses the effects of the smaller scales deformation at a given scale. This might yield less intermittency because the smallest scales which can lead to the largest exponentiated spikes will be suppressed. On the other hand, it may be an improvement for the coherence of the resulting structures compared to \citet{Chevillard2011}. Indeed, smoothing avoids strong deformations from very small scales which could result in less correlation between contiguous scales as the strongest peaks at the smallest scales might dominate every scale. In the case of the Zel'dovich Approximation, it was in fact recognised that some smoothing
improved the accuracy of the approximation for a given targeted evolution
time, which led to the truncated Zel'dovich Approximation \citep[e.g.,][]{ColesMelottShandarin1993}.

Future investigations of the MuScaTS method in the 3D incompressible HD case will decide which formulation performs best. 
Finally, note that MuScaTS also retains both advection (neglected in CRV) and diffusion (approximated by a regularisation cut-off at small scales in CRV).

\subsection{Links to the Zel'dovich multiscale approximation}
\label{sec:Zeldo}

In cosmology, the concept of multiscale is directly related to hierarchical formation of structures, in particular dark matter halos. In the concordance model of large-scale structure formation, dark matter halos are the hosts of galaxies and clusters of galaxies and their history consists of successive mergers. In this bottom-up scenario, small halos form first and then merge together to form a population of larger halos. The Lagrangian regions occupied by these halos define a smoothing scale, which can itself be associated with a mass and a time scale of halo formation \citep{PressSchechter1974}. In this picture, mergers are equivalent to connections between Lagrangian regions. As a result, the dynamical process at play can also be described with a multiscale approach accounting for dynamical collapse times of halos of various masses associated with a smoothing level. 

The Zel'dovich approximation (ZA) consists in a ballistic integration of initial velocities. Smoothing initial conditions to the ZA was shown to improve the solution, because it avoids the dispersion resulting from shell crossing. This led to the so-called truncated ZA \citep[e.g.,][]{ColesMelottShandarin1993} or higher order Lagrangian truncated perturbation theory \citep[e.g.][and references therein]{BernardeauCGS2002}, where the level of truncation is the smoothing scale. In the hierarchical formation of halos described above, the truncated ZA (or its higher order equivalent) can be improved further by a multiscale approach : a smoothing length depending on the local position is defined.  This multiscale synthesis of dark matter dynamics, has been implemented with success by different authors \citep[see, e.g.,][]{BondMyers1996,MonacoTTGQS2002,SteinAlvarezBond2019}. In these existing multiscale approaches, at every initial position, one finds the largest smoothing scale which does not collapse at the targeted redshift. This identifies both the smoothing length and the evolution time during which this Lagrangian region is evolved. 

As mentioned in section \ref{sec:scope-others}, the equations of large scale structure formation including the expansion of the Universe can be cast in our generic form \eqref{eq:generic-form}, and we use a ballistic approximation which in this context amounts to ZA. A multiscale ZA can therefore directly emerge from our MuScaTS framework. In our MuScaTS approach, we suggest to identify the coherence time with the \cite{PressSchechter1974} collapse time at a given smoothing length, capped by the total time corresponding to the targeted redshift. This results in two subtle differences compared to existing ZA approaches. First, the collapse time at an intermediate scale is computed on a field which has already been advanced at the targeted redshift on the largest scales. Second, every bandpass filtered density field is evolved separately during its own time scale. This may result in different performances compared to existing methods, and it remains to be checked whether they are better or worse. However, the principle of the resulting method is still very close to the original multiscale ZA approaches.

\section{Validation for 2D incompressible hydrodynamics}
\label{sec:validation}

In the previous sections, we presented our MuScaTS framework to generate random realisations of a variety of flows. We provided a possible numerical implementation of the procedure. We also connected our synthesis to previously known generating methods in a few specific contexts.

We now turn to validate the method in the simplest possible setup: incompressible 2D hydrodynamics. Our generative procedure deforms an initial Gaussian field under the influence of the flow. We will assess to what extent our synthesis is able to recover the statistics obtained when the flow evolves from an initial Gaussian field with a prescribed spectrum. 

To that end, we run accurate simulations of the flow (using a pseudo-spectral code  presented below in section \ref{sec:reference-simulations}) starting with a large number (here, 30) of independent initial random realisations of Gaussian fields with a prescribed power-law spectrum. 
As remarked in the presentation of our framework above, several methodological choices need to be defined. We consider a fiducial version of our synthesis method, \tcg{that we call the `basic' synthesis}, and introduce a few variants in section \ref{sec:variants} which span several ways of estimating the divorticity evolution and the coherence time. We quantitatively compare the performances of our syntheses first in the eyes of classical statistical indicators (section \ref{sec:classic}) and second with the help of texture sensitive statistics (section \ref{sec:WST}).

\subsection{Reference 2D simulations }
\label{sec:reference-simulations}
We briefly state here the details of our numerical implementation of a pseudo-spectral code to evolve 2D incompressible hydrodynamics. 

We integrate the evolution equation \eqref{eq:incompressible-2D} on a square periodic domain. We choose the length scale normalisation such that the domain size is $L=2\pi$  and define our velocity scale such that the {\it expectancy} of the initial velocity squared is unity, namely $E[\left<u^{2}\right>]=1$ where $E$ is the expectancy over all our realisations and $\left<.\right>$ denote the average over our computational domain. Note that we do not require that the initial r.m.s. velocity is unity for every realisation, as this would destroy the Gaussianity of our initial process. The Reynolds number is hence on average $R_{e}=E[\sqrt{<u^{2}>}]L/\nu=2\pi/\nu$ initially, and then decays as time proceeds while velocity is dissipated by viscous friction. Simulations are run with a pseudo-spectral method at a Courant number of 1, with a Runge-Kutta integration scheme of order 4, using a 2/3 truncation rule.

The initial field is chosen to be random, Gaussian with a prescribed Fourier spectrum $E_{u}(k)\propto k^{\beta}$. More specifically, we draw a random Gaussian white noise in real space, compute the Fourier coefficients, imprint the spectrum by multiplying the resulting coefficients by $C_\beta k^{\beta}$ where $C_\beta$ is chosen to comply with $E[\left<u^{2}\right>]=1$, and finally we get back to real space. The initial turnover time scale averaged over our realisations is hence $t_{\rm turnover time}=L/\sqrt{E[\left<u^{2}\right>]}=2\pi$ in our units.
The seed of the random noise is controlled by the integer parameter \texttt{seed} in the intrinsic FORTRAN random generator, which allows us to select various independent but deterministic realisations of the noise.

\begin{figure*}
    \centering
    \includegraphics[width=\linewidth]{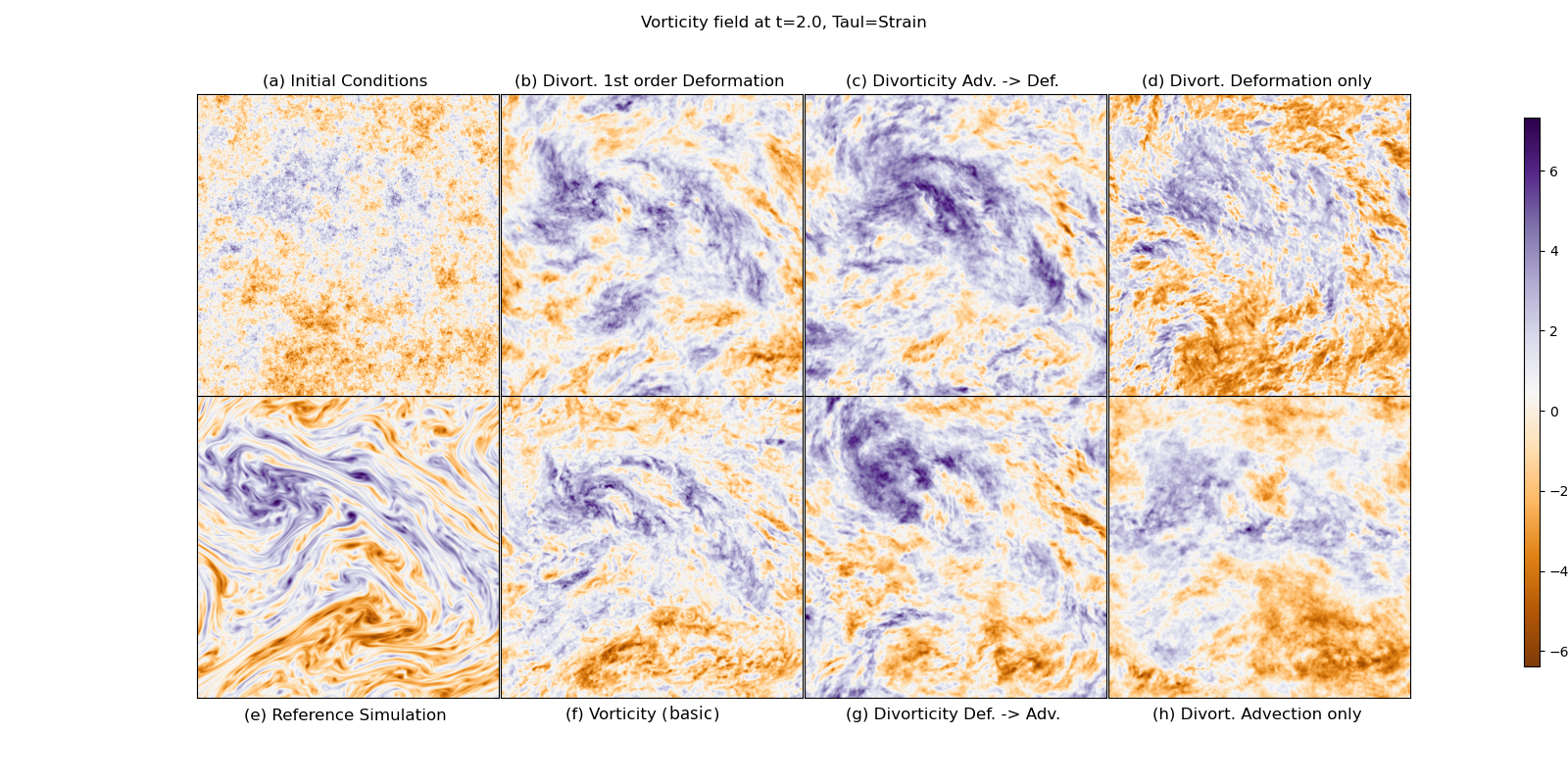}
    \caption{Vorticity maps for various cases (see text), all based on the {\it same} original Gaussian field (displayed on panel (a)). (a) initial conditions, (b-c-d) divorticity syntheses with (b) first order deformation (c) advection then deformation (d) deformation only, (e) reference simulation at $t=2$ (about 1/3rd turnover time scale) (f) basic synthesis for vorticity (g-h) divorticity syntheses with (g) deformation then advection (h) advection only. All syntheses are shown at the same time $t$ than for the simulation and a coherence time based on the local strain time (See subsection \ref{subsec:coherence-time}).}
    \label{fig:vorticity-maps}
\end{figure*}

The resolution is set by the number $N$ of grid elements per side of the square domain, hence the pixel size is simply $\Delta x=L/N=2\pi/N$.
The time step length is variable, defined by the Courant number parameter $\mathtt{CFL}$ through the relation $\Delta t=\mathtt{CFL}\Delta x/|u|_{max}$. We set the Courant number to $\mathtt{CFL}=1$. The top-left panel (a) on figure \ref{fig:vorticity-maps} illustrates one such initial vorticity map.

Our basic run is $N=1024$ pixels aside for a viscous coefficient $\nu=10^{-4}$ with an initial power-law spectrum with a slope $\beta=-3$. This power-law spectrum corresponds to the power-law slope predicted in stationary 2D turbulence for the enstrophy cascade from  the injection scale towards the small scales \citep{Kraichnan1967}.   

\begin{figure*}
    \centering
    \includegraphics[width=\linewidth]{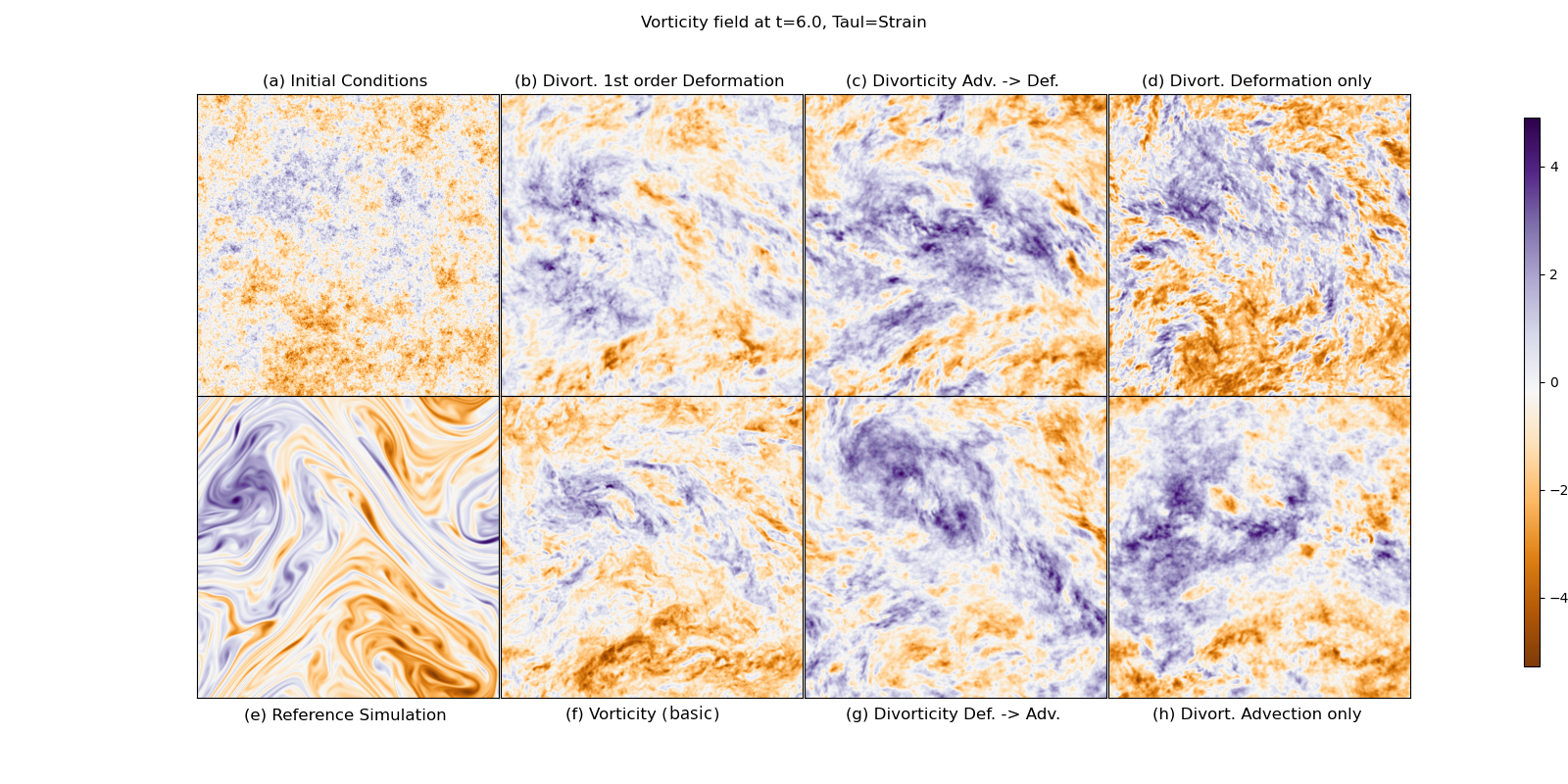}
    \caption{Same as figure \ref{fig:vorticity-maps}, but at time $t=6$, i.e. at about one turnover time scale.}
    \label{fig:maps-6}
\end{figure*}

We evolve the simulation from these initial conditions during a total integration time $t$. The bottom-left panel (e) on figure \ref{fig:vorticity-maps} illustrates the vorticity map after the initial Gaussian field has been evolved to $t=2$, i.e. during about a third of a turnover time scale. Figure \ref{fig:maps-6} present the same results after one turnover time-scale.
Clumps of vorticity quickly form, positive blobs rotate clockwise while negative ones (with $w<0$) rotate counter-clockwise, displaying distinctive trailing spiral arms. Fluid in between the vorticity clumps gets sheared into filaments extremely elongated after about one turnover time ($t=6)$.

\subsection{Basic synthesis and variants}
\label{sec:variants}

We compare the above reference simulation to several variants of our synthesis method. We first present the parameters for our basic synthesis. It advects vorticity, using a coherence time $\tau_\ell$ based on the local strain time and the target "age" $t$ of the simulation (see subsection \ref{subsec:coherence-time}). The filtering is realised with our cosine isotropic filter (see appendix \ref{sec:filters}). We use the same resolution $N$, spectral index $\beta$ and viscosity $\nu$ as for the simulations, and we compare results at the same global evolution time $t$ (see panels (e) and (f) in figure \ref{fig:vorticity-maps} for $t=2$ and figure \ref{fig:maps-6} at $t=6$). 

We now develop our investigation of the variants around our basic synthesis. In the remaining panels of figures \ref{fig:vorticity-maps} and \ref{fig:maps-6} (panels (b-d) and (g-h)), we consider several variants of a MuScaTS synthesis based on the divorticity field. In particular, we assess the relative importance of the advection and deformation (stretch in 2D) terms, both in order to prepare for the 3D case and in order to link to the previous works of CRV and \citet{Rosales2006}. We investigate the effect of the ordering of the advection term before or after the deformation term (see panels (c) and (g) of figure \ref{fig:vorticity-maps}). We investigate the effect of including only deformation as in CRV or only advection as in \cite{Rosales2006} (see panels (d) and (h) in figure \ref{fig:vorticity-maps}). Finally, we test the effect of computing the deformation at first order instead of developing it fully through a matrix exponential (panel (b) in figure \ref{fig:vorticity-maps}).

Eye inspection of figure \ref{fig:vorticity-maps} reveals that our basic synthesis (panel (f) for our synthesis on vorticity) does a rather good job at capturing the spiralling shape of the vorticity clumps, but the texture of the vorticity feels more patchy than in the simulation, specially at late times (see figure \ref{fig:maps-6}). The three syntheses based on divorticity which retain both advection and deformation (panels (b), (c) and (g)) seem to perform slightly less well, although the first order version of the deformation (panel (b)) is close to our basic synthesis (panel (f)). The two divorticity syntheses which skip one of these two terms (panel (d) and (h)) perform significantly less well. The stretch only case (panel (d)) recovers part of the filamentary texture and the advection only case (panel (h)) seems the worst, although the positions of the two large scale clumps of vorticity seem reproduced.

We also played with the coherence time $\tau_\ell$. We tested a constant $\tau_\ell$, which leads to a significant loss of realism (not shown here). And we tested a $\tau_\ell$ based on the local shell crossing time or the local stretch time (see subsection \ref{subsec:coherence-time}), without observing significant improvement from the strain based time. Finally, we tried both the Cosine and spline filters (see appendix \ref{sec:filters}) without observing much of a difference. 

Figures \ref{fig:vorticity-maps} and \ref{fig:maps-6} present a simulation and a few synthesis variants for a single {\it common} initial Gaussian field.
We will now assess the statistical quality of each variant of our synthesis by running a reference set of simulations with 30 independent initial Gaussian conditions. In the next sections, we will compute a selection of statistics on these 30 independent realisations to estimate their mean and variance. We will then compare the same statistics on each variant of our syntheses generated from 30 {\it other} independent Gaussian fields. We will thus get quantitative estimates of the performance of our synthesis for each choice of statistical indicators.

\subsection{Classical statistical indicators}
\label{sec:classic}

After \citet{K41} discovered his famous power-spectrum law for 3D incompressible hydrodynamics, \citet{Kraichnan1967} was the first to extend it to 2D incompressible hydrodynamics. He predicted that in forced 2D turbulence the velocity power spectrum should scale as $k^{-5/3}$ above the injection scale (indirect energy cascade) and as $k^{-3}$ below it (direct enstrophy cascade). This was later verified both in soap film experiments \citep{Rutgers1998,Bruneau2005} and in simulations \citep{Boffetta2007}.

Intermittent corrections to these power-law behaviours were shown to be very weak in the indirect cascade \citep{Boffetta2000}. Very weak non-Gaussianities were detected in the velocity increments probability distribution functions (PDFs), whether measured in experiments \citep{Paret1998} or simulations \citep{Boffetta2000,Pasquero2002}. The PDF of vorticity increments on the other hand were predicted to display exponential tails by \citet{Falkovich2011}. Indeed, significant deviations from Gaussianity were observed for the vorticity increments in simulations by \citet{Tan2014}. As a result, we choose to focus on vorticity increments rather than velocity increments since those will provide an easier measurement of intermittency compared to velocity increments.

Analogues of the exact K\'arm\'an Howarth Monin relation in 3D exist for both the indirect cascade of energy and the direct cascade of enstrophy in 2D.  \cite{Eyink1996} showed that in the indirect cascade, the energy transfer function
\begin{equation}
{\cal F}_u=\left<\delta_{\ell}u_{\parallel}^{3}\right>,
\end{equation}
where brackets denote both space and ensemble averaging, has to be positive and proportional to $\ell$. This is linked to slight asymmetries in the PDFs of the longitudinal velocity increments, with a bias towards positive values for $\delta_{\ell}u_{\parallel}$ at large values of $\delta_{\ell}u_{\parallel}$. On the other hand, in the direct cascade, the enstrophy transfer function 
\begin{equation}
{\cal F}_w=\left<\delta_{\ell}u_{\parallel}(\delta_{\ell}w)^{2}\right>
\end{equation}
was predicted to be negative and proportional to $\ell$ \citep{Paret1998}, see also \citet{Bernard1999}. This second result also involves asymmetries of the longitudinal velocity increments, but correlated to the vorticity increment (which is itself symmetric): it is caused by a bias for negative values of $\delta_{\ell}u_{\parallel}$ conditioned on large vorticity increments. In the following we will examine both the energy and enstrophy transfer functions ${\cal F}_u$ and ${\cal F}_w$ as probes of these second order behaviours of the increments PDFs.

The above results were obtained at steady state in driven turbulence. However, here we deal with decaying turbulence, so none of these results may hold. Nevertheless, we will use these classical indicators, power spectra, increments PDFs and transfer functions, as quantitative measurements to assess how our syntheses perform compared to actual simulations.

\begin{figure}
    \centering
    \includegraphics[width=1\linewidth]{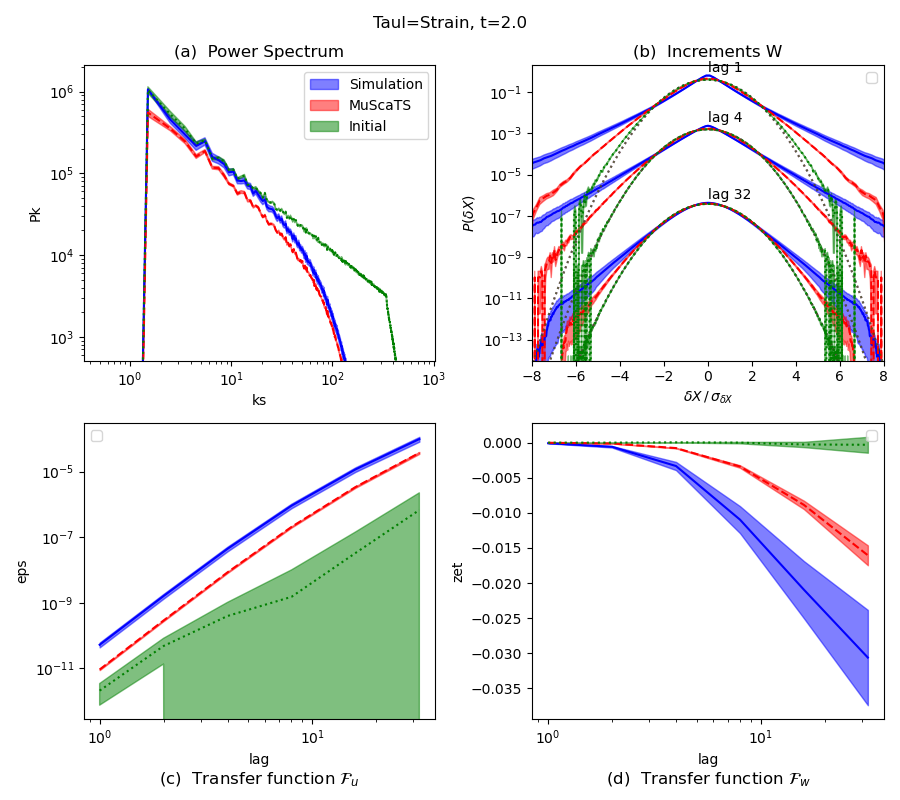}
    \caption{Classical statistics averaged over 30 realisations for (a) power spectrum, (b) vorticity increments (for lags of 1, 4 and 32 pixels), (c) energy transfer function ${\cal F}_u$ and (d) enstrophy transfer function ${\cal F}_w$. Green is for the initial conditions (of the syntheses), red for the MuScaTS syntheses (basic run) and blue the simulations, at time $t=2$. Shaded areas give the 1-sigma error around the mean over the 30 independent realisations (i.e. mean plus or minus standard deviation divided by $\sqrt{30}$).}
    \label{fig:classic}
\end{figure}

\begin{figure}
    \centering
    \includegraphics[width=1\linewidth]{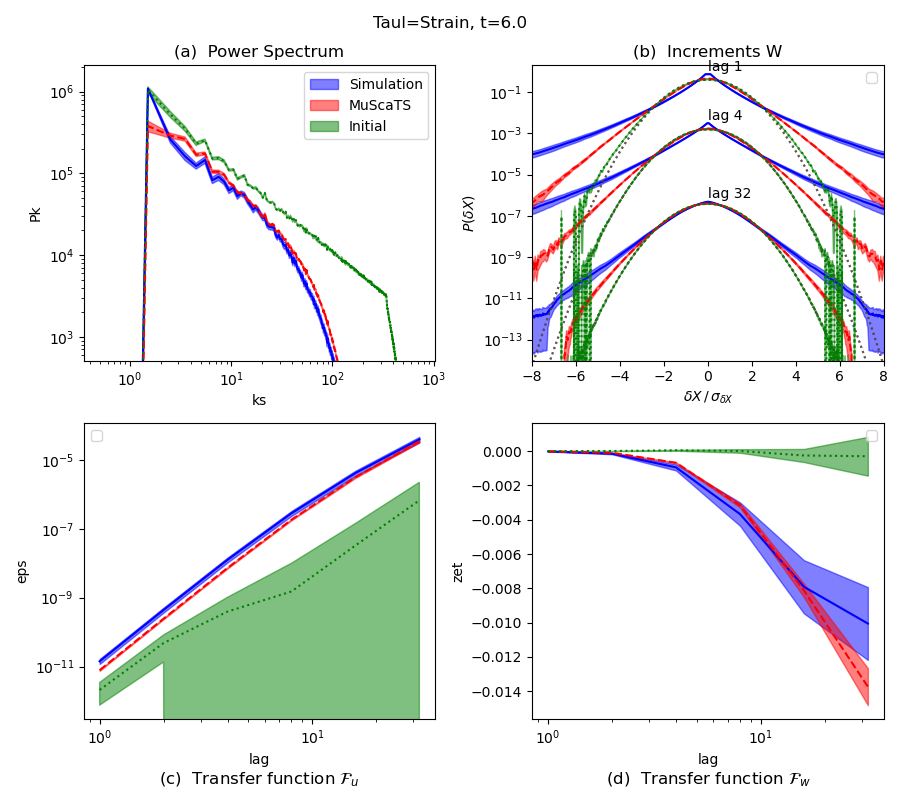}
    \caption{Same as figure \ref{fig:classic}, but at time $t=6$, i.e. at about one turnover time scale.}
    \label{fig:classic-6}
\end{figure}

\subsubsection{Power spectra}
\label{sec:power-spectra}
We compare in panel (a) of figure \ref{fig:classic} the average power spectra over all 30 realisations (reference simulations in blue and synthesis in red, while initial conditions are in green). Both the slope in the inertial range power-law behaviour and the power spectrum shape in the dissipation range seem equally well recovered. On these figures, the shaded area give the error on the mean as estimated from the dispersion over our 30 realisations.
It should be stressed that these error estimates are similar between the simulations and the synthesis. In other words, the synthesis is able to reproduce not only the average, but also the variance of the power spectrum across the 30 members of our statistical ensemble. After about one turnover time scale, both the power spectrum and its variance are still well accounted for (see figure \ref{fig:classic-6}).
 
\subsubsection{Increments statistics}

We compute $\delta_{\mathbf{\boldsymbol{\ell}}}X(\boldsymbol{r})=X(\boldsymbol{r}+\boldsymbol{\ell})-X(\boldsymbol{r})$ as the increment of a quantity $X$ over a lag $\boldsymbol{\ell}$ at position $\boldsymbol{r}$ in the map. We then examine the statistics of all increments which have their lag $\boldsymbol{\ell}$ in a corona of width 1 pixel around a prescribed norm in pixels. For instance, the PDF of vorticity increments at various lags (in pixels) is displayed in panel (b) of figure \ref{fig:classic}.
The PDF in the simulation experience large departures from Gaussianity. If we take as metric the logarithm of the PDF wings, at $t=2$ (i.e. a third of a turnover time), a significant fraction of the non-Gaussianity of vorticity increments is reproduced (synthesis takes us more than half way between Gaussian and initial conditions). Also, the relative dispersion of these results is well estimated, too. At one turnover time scale (see figure \ref{fig:classic-6}), the discrepancy between the PDFs from simulations and synthesis is larger (simulations have slightly increased their departure from Gaussianity while syntheses are still at about the same place).

\subsubsection{Transfer functions}

As mentioned above, in steady state forced 2D turbulence, the energy transfer function ${\cal F}_u$ is expected to be positive (indirect cascade) for scales above the forcing scale, while the enstrophy transfer function ${\cal F}_w$ should be negative for scales below. Although we have no forcing here, our decaying turbulence simulations display the same signs for the energy transfer functions. Early 3D generative models such as \cite{Rosales2006} or \cite{Chevillard2011} recognised that it is a real challenge to obtain random fields which can account for the non-zero transfer of energy. Here, our synthesis is able to account not only for the correct sign of these transports but also for their shape, as shown in panel (c) and (d) of figure \ref{fig:classic}. Their magnitude however is underestimated in the syntheses. The agreement is actually better at later times (see figure \ref{fig:classic-6}) because as time proceeds the total energy in the simulation decays while the syntheses become frozen as the strain time is reached everywhere.  Finally, most of the relative dispersion of these quantities is reproduced by the synthesis, at all times.

\subsection{Scattering transform statistics}
\label{sec:WST}
Scattering transform statistics are a modern and powerful tool characterising textures (Mallat 2012). In the context of cosmological simulations, \citet{2020PhRvD.102j3506A} have shown that wavelet phase harmonics are able to predict many other non-Gaussian indicators (increment statistics, Minkowsky functionals, bi-spectrum, etc\dots ). 

Here, we present an analysis of our results using wavelet scattering transform (WST) coefficients. The first layer WST coefficients of a field $X$ are defined as:
\begin{equation}
S_1(j)=\left<|X*\psi_{j,\theta}|\right>_\theta,
\end{equation}
where $\left<\right>_\theta$ denotes the combination of ensemble average, spatial average over the computational domain and average over $\theta$. They constitute a statistical probe similar to the power spectrum, with additional sensitivity to sparsity\footnote{The sparsity of a signal refers to its property to have a large fraction of its values being close to $0$. It can be measured by comparing its $L^1$ and $L^2$ norms. Such measure can be used to characterize the intermittent nature of turbulent flows~\citep{bruna2015intermittent}.} introduced by the use of the $L^1$-norm instead of the $L^2$-norm. We do not show the results, as they are very similar to those obtained for the power spectra (see section \ref{sec:power-spectra}).

\begin{figure*}
    \centering
    \includegraphics[width=\linewidth]{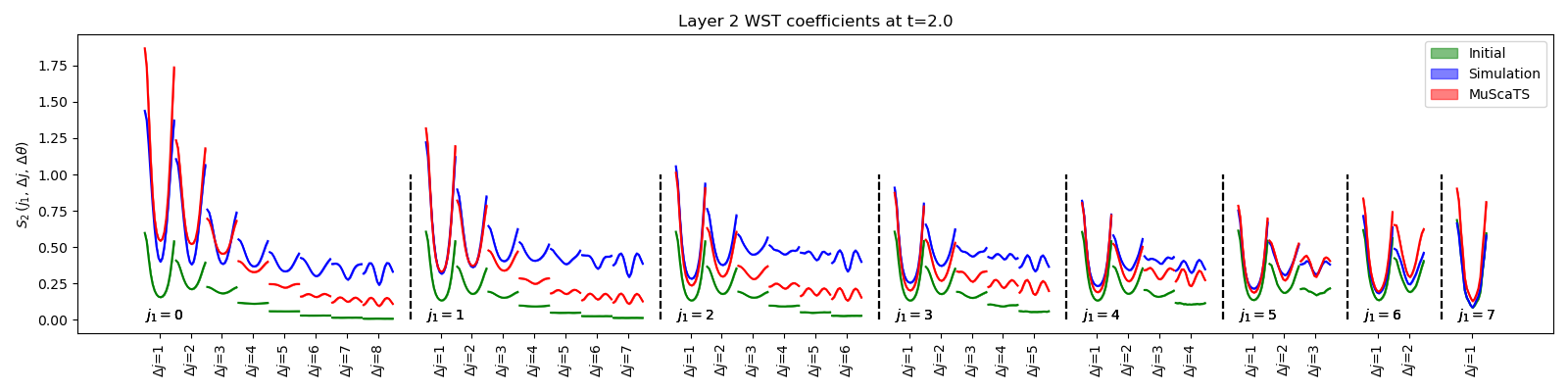}
    \caption{\label{fig:WST-coefficients} Mean over 30 realisations of layer 2 WST coefficients $S_2(j_1,j_2,\Delta \theta)$ (see text) for initial conditions (green), the reference simulation (blue) at about a third of a turnover time ($t=2$) and MuScaTS basic syntheses over a time $t=2$ (red). The main length scale runs from its index $j_1=0$ (smallest) to $j_1=J-3=7$. The secondary scale $j_2$ is labelled by $\Delta j=j_2-j_1$.  The relative angle $\Delta \theta$ runs from 0 to $\pi$ from left to right in each broken line of fixed $j_1$ and $\Delta j$. The width of the lines indicates the standard deviation of this mean over our 30 realisations, the computational domain and the absolute angle $\theta_1$. 
}
\end{figure*}

We then define normalised layer 2 coefficients as:
\begin{equation}
    S_2(j_1,j_2,\Delta \theta)= \frac{\left<||X*\psi_{j_1,\theta}|*\psi_{j_2,\theta+\Delta\theta}|\right>_\theta}
    {S_1(j_1) S_1(j_2)}.
\end{equation}
The layer 2 coefficients ($S_{2}$) estimate the coupling between two angles $\theta_{1}$ and $\theta_{2}=\theta_1+\Delta \theta$ and two scales $\ell_{1}=\ell_0 2^{j_{1}}$ and $\ell_{2}=\ell_0 2^{j_{2}}$ labelled by $j_{1}$ and $j_{2}$, with $\ell_0$ the smallest scale considered (note that by convention scales are labelled by \emph{increasing }order in WST coefficients).
In our case, thanks to isotropy, the coefficients depend only on the difference $\Delta\theta=\theta_{1}-\theta_{2}$ and we have checked that indeed there is very little dispersion of the coefficients at fixed $\Delta\theta$, $j_{1}$ and $j_{2}$ when $\theta_{1}$ varies. The shaded areas on figure \ref{fig:WST-coefficients} actually reflect the dispersion resulting from the spatial average over the vorticity fields, the ensemble average over our 30 realisations and from the angular average over $\theta_1$. These shaded area are so thin that they look like solid lines. With our 1024 pixels resolution, we chose 9 dyadic scales along with 16 angles. We plot on figure \ref{fig:WST-coefficients} the normalised  $S_{2}$ coefficients at $t=2$ (i.e. about one third of a turnover time).
 
 This powerful tool is much more stringent than the increments and probes much more accurately the defects of our synthesis. The synthesis correctly captures the relative variation of the coefficients with $\Delta\theta$ but fails at reproducing the proper slope of the power-law decay with $\Delta j=(j_{2}\lyxmathsym{\textendash}j_{1})$ (although the decay is still somewhat power-law like). At later times (not shown here), the angular variation is also lost by the synthesis. We stress here that although the classical indicators shown in the previous subsection \ref{sec:classic} were poorly sensitive to the texture differences, the WST coefficients provide a much better quantitative assessment of the discrepancies detected by an eye inspection of figure \ref{fig:vorticity-maps}. Indeed, WST statistics are known to produce remarkable syntheses~\citep{2019A&A...629A.115A} and constitute a very practical tool to partly assess quantitatively what our eyes are able to probe. First, in this metric, the distance between simulations and syntheses relative to the distance between initial conditions and final result is much larger compared to classical indicators. The synthesis works us a much smaller path towards the actual non-Gaussianities in the eyes of scattering statistics. Second, the discrepancy has a lot more statistical significance: the very small amount of dispersion for these layer 2 coefficients greatly helps disentangling the results between the simulations and our generative models.
 
  We also attempted to find good reduced models for the variation of the layer two coefficients with respect to the relative angle $\Delta \theta$, in the  spirit of the reduced WST presented in \cite{2019A&A...629A.115A}. We found that the generic shape $S_2(i,j,\Delta\theta)\propto \alpha_{i-j}+\beta_{i-j}\cos(2\Delta\theta)+\gamma_{i-j}\cos(4\Delta\theta)$ provides a good fit to the data, but opted not to show the resulting adjustments to prevent overloading the paper.

\begin{figure*}
    \centering
    \includegraphics[width=\linewidth]{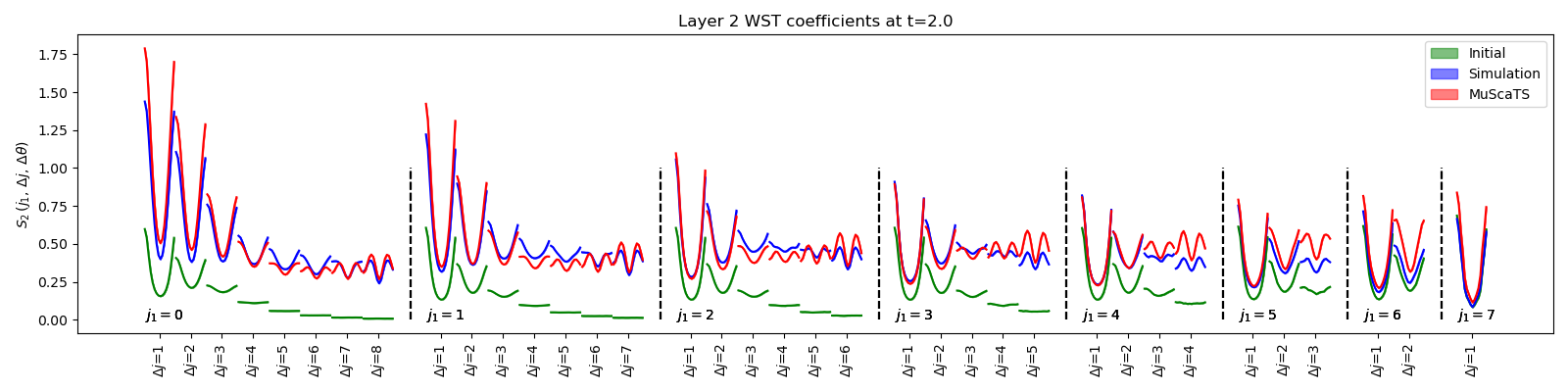}
    \caption{\label{fig:B4-WST-coefficients} Same as figure \ref{fig:WST-coefficients} but with a 4 steps bootstrap (see text of subsection \ref{subsec:bootstrap}, and see also figure \ref{fig:bootstrap}). 
}
\end{figure*}

\section {Discussion}
\label{sec:discussion}
We provide in this section a brief account of the parameter space investigation we have conducted, and discuss a few of the shortcomings of our method.

\subsection{Filter shape}

Although all the results of the present paper are computed with a bank of filters based on our cosine filters, we have also investigated B$_3$-spline filters (see appendix \ref{sec:filters}). There are very slight  differences between the two, hardly significant statistically and we do not show them here. We chose to present the cosine filters because they can cover the Fourier space at a slightly lower cost.
In a paper which inspired us for this work, \cite{Lewalle2010} showed how Gaussian based wavelets (Mexican hats) greatly simplify the analytical expressions for the filtered equations. Mexican hats might hence be best suited for theoretical work based on our MuScaTS framework (their weak selective power may hinder the quality of the syntheses, though). 

\subsection{Divorticity}
\label{subsec:divorticity}

As shown above (see section \ref{sec:generic-biv}), the evolution equations for 2D incompressible hydrodynamics for the divorticity also suit our generic formalism \eqref{eq:generic-form}, with both an advection and a deformation term present, very similar to the structure obtained for the 3D vorticity evolution equations. 

We find that our method performs very slightly better when we base it on the vorticity evolution equation rather than on the divorticity equation. This might be due of the presence of two terms (advection and deformation) in the divorticity evolution equation. Also, it may be due to the fact that the divorticity power spectrum is even more biased toward the small scales than the vorticity spectrum. Finally, the divorticity remapping introduces spurious non zero divergence in this divergence free field, while the vorticity field does not suffer from this, as it is a scalar in 2D.

We have investigated several variants in the implementation of our method applied to divorticity in 2D. We swapped the order between advection and deformation, without finding significant change (compare panels (c) and (g) on figure \ref{fig:vorticity-maps}). We tested the full matrix exponential implementation of equation \eqref{eq:advection-deformation} and its linearised version in $\tau_\ell$, with a slight preference for the latter implementation (compare panel (b) and (c) on figure \ref{fig:vorticity-maps}). This might be due to the better consistency between the ballistic approximation for the advection term and a first order approximation in time for the deformation. Classical indicators are almost blind to whether vorticity or divorticity is advanced, with a slight preference for the latter method for the vorticity increments while transfer functions are better reproduced in the former case. Vorticity maps and WST coefficients give a slight preference to the treatment of vorticity rather than divorticity.

The above choices retaining both advection and deformation for the implementation of MuScaTS on the divorticity were not critical. But we found a huge difference when either the advection or the deformation term were omitted, as panels (d) and (h) in figure \ref{fig:vorticity-maps} can testify. Even the classical indicators suffer from the loss of either of these terms. This is a good sign for future implementation of our method in the 3D case which should perform even better when compared to existing methods such as \cite{Rosales2006} or \cite{Chevillard2011} and their extensions.  

\subsection{Choosing the coherence time}

Most results presented here have used a coherence time based on a smooth minimum between the local strain time (see our last choice in section \ref{subsec:coherence-time}) and the global target time. However, we have also investigated the usage of a constant global target time, with significantly less accurate results. We also tested a coherence time uniform in space, but dependent on scale as a power-law (which resulted in significantly less accuracy). And finally we tested the other options based on the local shell crossing time or the local stretch time and found the method appeared slightly less accurate in view of the statistical indicators we tested.
Perhaps future investigations in compressible or 3D applications will help us better understand which version of the coherence time is most appropriate.

\subsection{Bootstrapping}
\label{subsec:bootstrap}
\begin{figure}
    \centering
    \includegraphics[width=1\columnwidth]{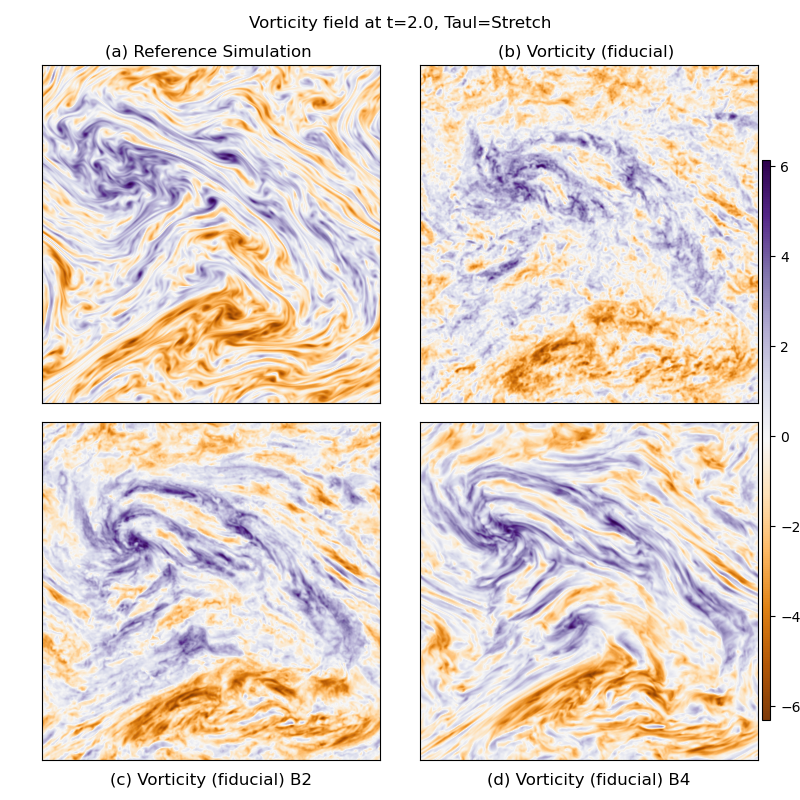}
    \caption{Vorticity maps for one realisation of our basic case at $t=2$ for (a) the simulation (b) same as panel (d) of figure \ref{fig:vorticity-maps} (c) a two-steps ($\Delta t=1)$ bootstrap and (d) a four steps ($\Delta t=0.5)$ bootstrap (see text).}
    \label{fig:bootstrap}
\end{figure}

Since our method performs better during a smaller fractions of the initial turnover time scale, it is natural to try and start a new MuScaTS step from the resulting snapshot obtained instead of using the initial Gaussian conditions. We have hence split the integration step in two or four smaller successive MuScaTS steps to test the convergence of the method, a process we denote as bootstrapping. The resulting improvement on the vorticity maps is striking, as illustrated on figure \ref{fig:bootstrap}.

Bootstrapping is a poor man's way of probing higher order schemes, hence it augurs well for the performance of future higher order methods. It is also a sign that for steady state turbulence we need better statistics than Gaussian conditions to randomly reshuffle our stochastic terms: including some non-Gaussianity (that of the previous step, when bootstrapping) in these pseudo initial conditions makes an important improvement. This resonates with the findings of \cite{Luebke2024} who found realistic coherent MHD structures when they started the method of \citet{Subedi2014} with a non Gaussian initial field.
Finally, the realism of the resulting solution shows that despite the strong approximations we made in deriving our filtered equations, we did not lose too much of the essential physics.

\subsection{Power spectrum shape}

All results shown in this paper are based on an initial velocity power spectrum proportional to $k^{-3}$. However, we have also tested a less steep power spectrum, scaling as $k^{-5/3}$, corresponding to the indirect energy cascade. This last spectrum is less dominated by the large scales, and in particular the vorticity field and more importantly the stretching field is then dominated by the small scales. This may explain why we have found our statistical comparison is significantly less good in this shallower spectrum case (not shown here). 

Indeed, in the current application of our method we apply a single sweep through scales from the largest to the smallest, implicitly considering scale causality in a unique direction. In future work, we will investigate the possibility to implement corrections to account for the feedback of the small scales to the largest (see further discussion in subsection \ref{subsec:small-terms}). 

In this paper, we have investigated only decaying turbulence. In principle, it is easy to incorporate a forcing term in our partial differential equations. However, a shortcoming of our method is that we have to introduce a power spectrum for the "initial conditions" $\boldsymbol{W}_{\ell}^{0}$. To estimate this power spectrum, we could resort to theoretical inputs, we could bootstrap our method until we get a steady power spectrum, or there may exist analytical considerations on energy conservation to provide the power spectrum shape from the filtered equations alone.

\subsection{Small-scale terms}
\label{subsec:small-terms}

We have currently bundled all small scale terms of our scale filtered equation \eqref{eq:full-filtered} into a "stochastic small scales" term with which we deal with in an extremely expedient manner. A large legacy of literature tackles coarse-graining or Reynolds averaging, which leads to convective terms, turbulent diffusion, small scale dynamo, eddy viscosity which all result from the filtering of the small scales terms. Estimation of these terms requires some sort of closure. Most methods adopt some sort of linear approximation to compute correlations between small scale terms, such as first order smoothing approximation (FOSA), also known as second order correlation approximation (SOCA), as introduced in \cite{Krause1980}, or mixing length theory approaches \citep[MLT, see][and references therein]{Lesaffre2013,Jermyn2018}. Test-field techniques have sometimes been used to validate these approaches in numerical simulations \citep{Brandenburg1998,Kapyla2020}.  

Here, we simply have short-circuited these considerations by randomly reshuffling the state variables of the gas over one coherence time-scale. 
However, once we have a first guess for the properties of our flow thanks to a first sweep from large to small scales, we could imagine refining the scale filtered evolution equations thanks to these terms which can be readily computed on that first guess. This would recover some of the causal influence of the small scales onto the large ones that our method currently completely neglects. 

It seems paradoxical that we can still get correlations such as energy transfers from small to large scales (as in our 2D example above where we get a positive energy transfer function). This means that in our 2D HD synthesis, large scales conspire to arrange velocities in such a way that there is a net energy transfer towards larger scales. Only, due to the single direction of our sweep from large to small scales, that energy transfer is not imprinted on the evolution equations. In our current investigation, it does not seem to be critical. However, there may be applications where it could be important, in particular when the spectrum is shallower, but also in cases when small-scales dynamos are at play (i.e. when some small scales instabilities lead to exponential growth of larger scale averaged quantities).

We finally point here to a slight inconsistency in our formalism: the filtered field $\boldsymbol{\tilde{W}}_{\ell}$ at scale $\ell$ may acquire power at other scales from its approximate filtered evolution equation \eqref{eq:Wtildel-vtilde}. This may happen through convergence and divergence of trajectories in the ballistic step, for example, but any non linear term will generate scales outside the initial range of the filter at scale $\ell$. At this stage, we are not sure whether it is a potential drawback or an actual advantage of our method (smaller scales build more efficiently, and in a natural way). In practice, however, we checked that the power generated at a scale $\ell$ did not overlap too much on neighbouring scales.  

\subsection{Extending the scope of our method}
  
We delve back here on the scope of our method. The generic form of the evolution equations \eqref{eq:generic-form} allows already for quite a few situations which were not dealt with before with generative methods. We chose this form because it behaves well under filtering and because it readily expresses advection and deformation, which clearly are useful concepts in fluid mechanics. This also allows us to easily link to previous works which had separately focused on either one of these two terms. However, it may not be the most general form which is amenable to our hierarchical filtering. To start with, we could accept various components of $\boldsymbol{W}$ to be advected individually with different speeds (as in the Elsasser formulation of incompressible MHD where $Z^+$ is advected by $Z^-$ and conversely) as long as these speeds remain linear in the state vector $\boldsymbol{W}$. There may also exist more general forms which survive filtering and the conservative form is certainly one which should attract our attention.
  
In addition, if we now stick to our generic form \eqref{eq:generic-form}, it is also not always easy to transform common partial differential equations towards it. For instance in the simple case of incompressible HD, we could not find a way to write down the velocity evolution equation (Navier-Stokes) in this form (there might be one that we could not see, or there might be a way to still apply the rest of the formalism to it, we simply have not found it yet). 

Finally, even in cases where we could not strictly fit an evolution equation to our framework, as in the case of compressible isothermal MHD, it may be that hiding isolated problematic terms into our stochastic reshuffling may not be too detrimental to the accuracy of our method.
 
\section{Summary and prospects}

We have investigated a set of few simple ideas applied to a large class of partial differential equations: hierarchical smoothing, first order evolution from random intermediate initial conditions during a scale and space dependent coherence time. With this procedure, we efficiently generate fields with realistic statistical properties compared to the results of the full time integration, at least in the simplest case of 2D incompressible HD.

Previous generative methods applied to 3D HD or MHD have focused either on advection \citep{Rosales2006,Subedi2014,Luebke2024} or on deformation \citep{Chevillard2011,Durrive2020,Durrive2022}. The present approach retains every physical term in the equations, including diffusion, and provides a general framework suitable to a large class of partial differential equations. In the 2D incompressible HD case, for the evolution of divorticity, we have shown that it is critical to include both advection and deformation at the same time: this brings hope that our method when applied to 3D HD or MHD may significantly outperform those existing techniques.

Classical statistical indicators remain relatively well reproduced by our approximations during a significant fraction of the turnover time (about one third). A few coherent structures such as vorticity clumps with spiralling arms are also reproduced, with the proper winding sense of rotation according to the vorticity sign. However, more stringent probes of the texture such as scattering statistics are not fooled and show discrepancy between simulations and syntheses. Yet a limited amount of bootstrap is able to bridge that gap: higher order integration schemes are likely to yield syntheses of much better quality at a smaller computational cost than bootstrap.

With the advent of deep learning, very efficient generative methods relying on deep learning architectures have emerged, as for instance diffusion models \citep[see][and references therein]{Genuist2024}. However, these techniques suffer from two strong shortcomings: the generative procedure is often difficult to apprehend, and usually needs a training set of considerable volume, which in this context means that a lot of full fledged numerical simulations have to be run prior to training the generative model. Our method, on the contrary, is directly fed with the physics contained in the partial differential equations and provides generated fields at the cost of a handful of simulation steps. Incidentally, our method could potentially be used to build large approximate training sets instead of resorting to actual simulations, or to complement a smaller set of accurate simulations with a large number of MuScaTS generated snapshots. Other generative methods as maximum entropy models conditioned by scattering transform statistics, are also being developed~\citep{bruna2019multiscale, zhang2021maximum, cheng2024scattering}, which require a very small training set (only a handful simulations and sometimes even a single snapshot). However, these methods generate a significant extra numerical cost and offer less control of the distance to the physical solutions compared to the method we propose.

The original purpose of the synthesis methods as setup by \cite{Rosales2006} was to reduce the time spent by simulations in the initial transient phase from arbitrary initial conditions towards fully developed turbulence. Incidentally, this reduces the carbon footprint of such simulations. Of course, the present model can still be used to the same aim. Here, we propose a range of other future applications. The low generating cost could be used with Monte Carlo Markov Chain techniques to adjust a synthetic field to some observational constraints. For example, one could imagine generating 3D synthetic fields with prescribed projections: to what extent can we recover the underlying 3D structure behind plane-of-sky astrophysical images in such a way ? Similar methods may be used to guess a turbulent field outside regions where it is measured, which could be useful in weather forecasting applications where reliable measurements are sometimes sparsely located. Finally, we hope our simplified models can be used as theoretical basis to understand existing or predict new statistical laws of turbulent flows.

\bibliographystyle{aa}
\bibliography{synthetic}

\section*{Acknowledgements}

We express gratitude to the referee for having read our paper so
carefully and for his constructive comments.
The authors acknowledge Interstellar Institute's programs "With Two Eyes" and "II6" and the Paris-Saclay University's Institut Pascal for hosting discussions that nourished the development of the ideas behind this work. PL acknowledges enlightening discussions with Marie Farge who accepted to read through an early version of the paper. PL acknowledges support from the European Research Council, under the European Community’s Seventh framework Programme, through the Advanced Grant MIST (FP7/2017-2022, No 742719). Computations were carried out on the \texttt{totoro} machine financed by the same ERC grant and hosted at the meso-center mesoPSL.

\appendix

\section{Reconstruction formula}
\label{sec:reconstruction_formula_demo}

In this appendix, we justify the reconstruction formula \eqref{eq:reconstruction}.

Let us take the Fourier transform of the right hand side of \eqref{eq:reconstruction}, considering it at a fixed given non zero $\boldsymbol{k}$. By linearity of the Fourier transform (interchanging the integral of the Fourier definition with the integral over scales $\ell$),
\begin{equation}
\label{eq:reconstruction_formula_demo_step1}
    \widehat{\left(\int_{-\infty}^{+\infty}{\rm d}\ln \ell \, f_{\ell}\right)}(\boldsymbol{k})
    =\int_{-\infty}^{+\infty}{\rm d}\ln \ell \, \hat{f_{\ell}}(\boldsymbol{k}).
\end{equation}
Moreover, by the convolution theorem (the Fourier transform of the convolution of two functions is the product of their Fourier transforms) relation \eqref{eq:filtering} in Fourier space, combined with \eqref{eq:dilation-k} and \eqref{eq:isotropic_filter}, lets us rewrite the integrand on the right hand side of \eqref{eq:reconstruction_formula_demo_step1} as
\begin{equation}
    \hat{f_{\ell}}(\boldsymbol{k}) = \hat{f}(\boldsymbol{k}) \hat{\varphi_{\ell}}(\boldsymbol{k}) = \hat{f}(\boldsymbol{k}) \Phi( |\boldsymbol{k}|\ell).
\end{equation}
Then, as $\hat{f}(\boldsymbol{k})$ is independent of $\ell$ it can be pulled out of the integral, and using the change of variable $s=|\boldsymbol{k}|\ell$ (possible since $\boldsymbol{k}$ is fixed and assumed to be non zero), we get that
\begin{equation}
\label{eq:reconstruction_formula_demo_step2}
\widehat{\left(\int_{-\infty}^{+\infty}{\rm d}\ln \ell \, f_{\ell}\right)}(\boldsymbol{k}) =
\hat{f}(\boldsymbol{k})\int_{-\infty}^{+\infty}{\rm d}\ln s \, \Phi(s)=C_\Phi\hat{f}(\boldsymbol{k}),
\end{equation}
where definition \eqref{eq:normalisation-constant_def2} has been used in the last step.
Given our normalisation \eqref{eq:normalisation}, relation \eqref{eq:reconstruction_formula_demo_step2} demonstrates that both sides of \eqref{eq:reconstruction} have the same Fourier transform for all non zero $\boldsymbol{k}$. For the special $\boldsymbol{k}=\boldsymbol{0}$ case, this statement remains true provided $\hat{f}(\boldsymbol{0})=0$ (i.e. $f$ has a zero mean) because all filters have zero means. Finally, by inverse Fourier transform, we obtain the reconstruction formula \eqref{eq:reconstruction}.

\section{Two example filters}
\label{sec:filters}
In the present application, we tested two sets of filters. The first one is based on  a cosine window in the logarithm of the scale. We first define the bump-like function

\begin{equation}
C_{2}(s)=\cos^{2}\left(\frac{\pi}{2}s\right),
\end{equation}
for $s$ in $[-1,1]$, and zero elsewhere. We then define our list
of filters as

\begin{equation}
\hat{\varphi}_{j}^{C}(\boldsymbol{k})=C_{2}(\ln[|\boldsymbol{k}|/k_{j}]/\ln[\lambda]),
\end{equation}
where $\lambda$ is the scale ratio between two adjacent scales. Our normalisation condition requires $J=J_C=\lfloor\ln(N/2)/\ln\lambda\rfloor$ (where $\lfloor\cdots\rfloor$ denotes the integral part or floor function) in order to have $\sum_{j=0}^{J}\hat{\varphi}_{j}^{C}(\boldsymbol{k})=1$ for all wavenumbers in the computational domain. Note that the maximum wave-number represented in our Fourier space is $k_{{\rm max}}/k_{J}=\sqrt{3}N/2$ for a 3D grid, or $k_{{\rm max}}/k_{J}=\sqrt{2}N/2$ for a 2D grid, and since the non-linearities in our method are quadratic, we use the two-third truncation rule in Fourier.

Our second set of filters is based on the box spline of 3rd order (often used in astrophysical applications due to its implementation efficiency, see appendix A of \citealp{2002aida.book.....S}, for example). Consider the $B_{3}$-spline cubic function (a box convolved three times with itself)

\begin{equation}
B_{3}(s)=\frac{1}{12}(|s-2|^{3}-4|s-1|^{3}+6|s|^{3}-4|s+1|^{3}+|s+2|^{3}).
\end{equation}

$B_{3}(s)$ is non zero only within the interval $[-2,2]$. It makes a smooth transition between 2/3 at $s=0$ and 0 at $s=2.$ We can use this smooth step property to build a smooth logarithmic window:

\begin{equation}
\hat{\varphi}_{j}^{S}(k)=\frac{3}{2}\left[B_{3}(|k|/k_{j+1})-B_{3}(|k|/k_{j})\right],
\end{equation}
where the $3/2$ coefficient is chosen to comply with our normalisation condition. The sum  $\sum_{j=0}^{J}\hat{\varphi}_{j}^{S}=\frac{3}{2}B_{3}(|k|/k_{J+1})$ has limit 1 when $|k|/k_{J+1}$ goes to inifinity, i.e. when $J$ is large enough for $k_{J+1}$ to be well above the maximum effective $k$ reached by the wavenumbers ($k_{{\rm max}}/k_{J}=\sqrt{3}N/2$ for a 3D grid, or $k_{{\rm max}}/k_{J}=\sqrt{2}N/2$ for a 2D grid). The normalisation condition is hence only approximately realised in this case.

\begin{figure}
\includegraphics[width=1.\columnwidth]{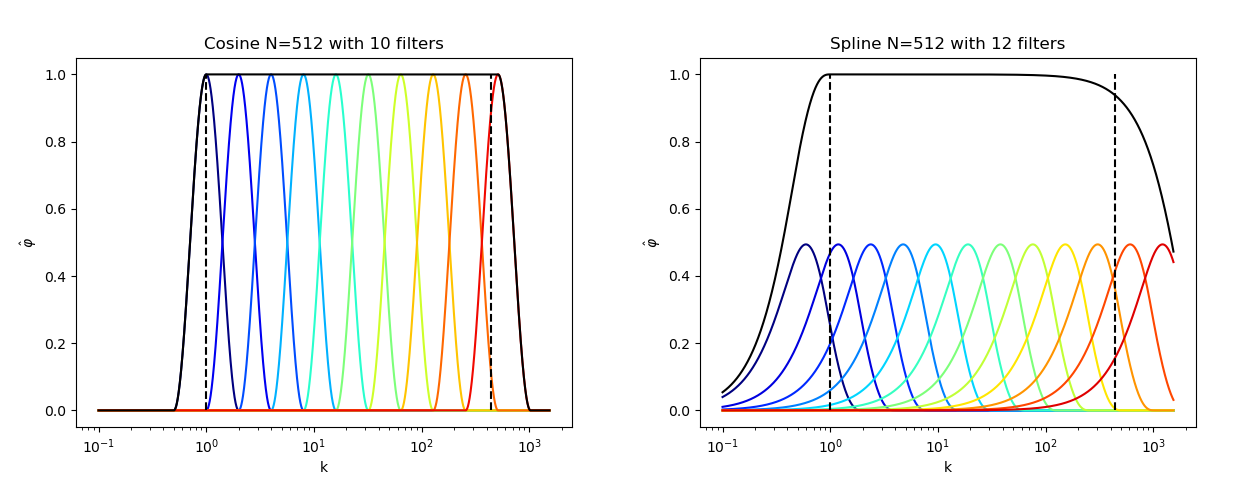}

\caption{\label{fig:Filters}Series of cosine (left) and  box spline (right)
filters for a cubic periodic domain of size $N=512$ pixels. Their total
sum is indicated as well as the wavenumber boundaries for a 3D domain
($k_{{\rm min}}=1$, $k_{{\rm max}}=\sqrt{3}N/2$). Colours from red to
blue stand for filter indices 0 to $J$. This figure illustrates how many filters
are necessary in order for the reconstruction formula to be valid.
In practice however, diffusion damps the higher $k$ end of the spectrum,
and convergence is reached for a slightly lower number of filters.}
\end{figure}

Figure \ref{fig:Filters} displays series of filters for these two choices of mother filter, along with the normalisation condition. The cosine filters are slightly narrower in scales compared to the splines. The spline filters do not match the condition exactly but only to the limit where we take a large number of filters. However, in practice, only the small scales will suffer from lack of power, which is in line with diffusion processes anyway, so it turns out that the use of $J_S=J_C+1=\lfloor\ln(N/2)/\ln\lambda\rfloor+1$ for the spline filters appears quite well converged, with only about 14\% of the total power missing compared to the limiting case (estimated from using $J=J_C+10$), and only 1\% missing at $J=J_C+2$, which we adopted in the present work.

\end{document}